\documentclass[prd,amsmath,amsfonts,floatfix,nofootinbib,preprintnumbers,superscriptaddress,showpacs]{revtex4}

\usepackage{graphicx}

\def\op{{\mathcal{O}}}
\def\etaprime{\eta^{\prime}}
\def\TrD{{\mathrm{tr}}_{{\mathrm{D}}}}
\def\vslash{v\!\!\!/}

\begin{document}
%%%%%%%%%%% Titlepage

\preprint{DAMTP-2006-123}
\preprint{NT@UW-06-32}

\title{Matrix elements of the complete set of $\Delta B = 2$ and
  $\Delta C = 2$ operators \\ in heavy meson chiral perturbation
  theory}

\author{William Detmold} \affiliation{Department of Physics,
  University of Washington, Box 351560, Seattle, WA 98195, U.S.A.}

\author{C.-J. David Lin}
 \affiliation{DAMTP, CMS, University of Cambridge, 
  Wilberforce Road, Cambridge CB3 0WA, England}
  \affiliation{Physics Division, 
  National Centre for Theoretical Sciences,
  Hsinchu 300, Taiwan}

\begin{abstract}
  Using heavy meson chiral perturbation theory, we consider the light
  quark-mass and spatial volume dependence of the matrix elements of
  $\Delta B=2$ and $\Delta C=2$ four-quark operators relevant for
  $B^{0}_{(s)}{-}\bar{B}^{0}_{(s)}$ and
  $D^{0}{-}\bar{D}^{0}$ mixing, and the $B_{s}$ meson width
  difference.  Our results for these matrix elements are obtained in the
  $N_{f}=2+1$ partially quenched theory, which becomes full QCD in the
  limit where sea and valence quark masses become equal.  They can be
  used in extrapolation of lattice calculations of these matrix
  elements to the physical light quark masses and to infinite volume.
  An important conclusion of this paper is that the chiral
  extrapolations for matrix elements of heavy-light meson mixing beyond
  the Standard Model, and those relevant for the $B_s$ width
  difference are more complicated than that for the Standard Model
  mixing matrix elements.
\end{abstract}
\pacs{11.15.Ha,12.38.Gc,12.15Ff}

\date\today \maketitle

\section{Introduction}
\label{sec:intro}

Neutral heavy-light meson mixing systems play a crucial role in
precision tests of the Standard Model and the search for new physics.
With the recently measured $\Delta m_{s}$ \cite{Abulencia:2006ze},
we can hope to obtain stringent constraints on the unitarity triangle of
the Cabibbo-Kobayashi-Maskawa (CKM) matrix, provided that the hadronic
matrix elements of the $B^{0}{-}\bar{B}^0$ and
$B_{s}^{0}{-}\bar{B}_s^0$ mixing processes are reliably
calculated.  On the other hand, the $D^{0}{-}\bar{D}^0$ mixing
system is a good channel to search for new physics
\cite{Bergmann:2000id}, because the Standard Model contribution is
strongly suppressed.

In the Standard Model, the short distance contribution to
the mass differences of the heavy neutral meson mixing systems
($B^{0}{-}\bar{B}^0$, $B_{s}^{0}{-}\bar{B}_s^0$ and
$D^{0}{-}\bar{D}^0$) is predominantly determined by the matrix
elements of a single set of four quark operators:
\begin{eqnarray}
  \label{eq:1}
   {\cal O}_{1,aa} &=& \bar{h}^{\alpha} \gamma_{\mu} (1-\gamma_{5})
   q_{a}^{\alpha} 
           \mbox{ }\bar{h}^{\beta} \gamma_{\mu} (1-\gamma_{5}) q_{a}^{\beta} ,
\end{eqnarray}
where $h$ is a heavy quark field (either a $b$ or a $c$ quark),
$q_{a}$ is a light-quark field with flavour $a$ ($a$ is not summed
over), and $\alpha$ and $\beta$ are colour indices.  Models containing
flavour-changing currents other than the $V-A$ form (arising in
supersymmetric extensions of the Standard Model and other scenarios)
usually result in mass differences that additionally depend on matrix
elements of the four-quark operators~\cite{Gabbiani:1996hi}
\begin{eqnarray}
 {\cal O}_{2,aa} &=& \bar{h}^{\alpha} (1-\gamma_{5}) q_{a}^{\alpha}
           \mbox{ }\bar{h}^{\beta} (1-\gamma_{5}) q_{a}^{\beta} ,
\nonumber\\
\label{eq:quark_level_op_susy} 
 {\cal O}_{3,aa} &=& \bar{h}^{\alpha} (1-\gamma_{5}) q_{a}^{\beta}
           \mbox{ }\bar{h}^{ \beta} (1-\gamma_{5}) q_{a}^{\alpha} ,
\\
 {\cal O}_{4,aa} &=& \bar{h}^{\alpha}  (1-\gamma_{5}) q_{a}^{\alpha}
           \mbox{ }\bar{h}^{\beta}   (1+\gamma_{5}) q_{a}^{\beta} ,
\nonumber\\
 {\cal O}_{5,aa} &=& \bar{h}^{\alpha}  (1-\gamma_{5}) q_{a}^{\beta}
           \mbox{ }\bar{h}^{\beta}  (1+\gamma_{5}) q_{a}^{\alpha} ,
\nonumber
\end{eqnarray}
(the right-handed analogues of $\op_{i,aa}$ for $i=1,2,3$ can also
contribute but their matrix elements are the same as those above as
the strong-interaction conserves parity).  Generically we can
represent these operators as
\begin{eqnarray}
  \label{eq:2}
  \op_{i,aa}= \bar{h} \Gamma_1 q \, \bar{h} \Gamma_2 q \,,
\end{eqnarray}
for the appropriate choice of spin and colour matrices,
$\Gamma_{1,2}$.  In lattice calculations, it is convenient to perform
a Fierz transformation which renders linear combinations of the
operators in Eq.~(\ref{eq:quark_level_op_susy}) into products of
colour-singlet currents. We choose to work in the basis of
Eq.~(\ref{eq:quark_level_op_susy}).

A subset of the operators in Eqs.~(\ref{eq:1}) and
(\ref{eq:quark_level_op_susy}) are also relevant for calculation of
the width-difference in the $B_s$ system, $\Delta
\Gamma_{B_s}/\Gamma_{B_s}$.  This difference is the largest amongst
the beauty hadrons ($\Delta \Gamma_{B_s}/\Gamma_{B_s} =
0.31{\tiny\begin{array}{c} +0.11\\-0.13
  \end{array}}$ \cite{Yao:2006px}) and,
following an operator product expansion, is given by
\cite{Beneke:1996gn},
\begin{eqnarray}
  \label{eq:3}
  \frac{\Delta \Gamma_{B_s}}{\Gamma_{B_s}}&=&\frac{G_F^2 m_b^2}{12\pi
    M_{B_s}} |V_{cb}V_{cs}|^2\tau_{B_s}
\left[ G(z) \langle \bar{B}_s^0|\op_{1,ss}|B_s^0\rangle +
 G_S(z) \langle \bar{B}_s^0|\op_{2,ss}|B_s^0\rangle \right]
+\op(1/M_b) ,
\end{eqnarray}
where the functions $G(z)$ and $G_S(z)$ are known at
NLO in perturbative QCD~\cite{Beneke:1998sy}. At ${\cal O}(1/M_b)$, matrix elements of
${\cal O}_{3,ss}$ also enter \cite{Lenz:2006hd}.

Lattice QCD is the only method for calculating the
$B^{0}{-}\bar{B}^{0}$ and $D^{0}{-}\bar{D}^{0}$ matrix elements
of the operators in Eqs.~(\ref{eq:1}) and
(\ref{eq:quark_level_op_susy}) from first principles, and much effort
has gone into such calculations (see Ref.~\cite{Onogi:2006km} for a
recent review).  However the existing lattice calculations have been
performed at light quark masses significantly larger than the physical
values and necessarily in finite volumes.  The effects of these
approximations need to be understood.  In this paper we consider the
light-quark mass extrapolation to the physical values for the lattice
calculations of these matrix elements.  Our framework is heavy meson
chiral perturbation theory (HM$\chi$PT) \cite{Burdman:1992gh,
  Wise:1992hn,Yan:1992gz} at finite volume \cite{Arndt:2004bg}. The
standard model $\Delta B=2$ operator $\op_{1,aa}$ has been considered
in this context in Ref.~\cite{Arndt:2004bg} and here we extend that
analysis to the full set of operators discussed above. As appropriate
for current and foreseeable lattice calculations, we work in the
isospin limit of SU(3) heavy meson chiral perturbation theory and give
results in the SU(6$|$3) partially-quenched extension. Primarily, we
treat the heavy quark as static throughout but consider the leading
effects of the splitting between the heavy-light vector and
pseudo-scalar mesons.  

An important conclusion of this work is that the chiral extrapolation
for matrix elements of ${\cal O}_{1,aa}$ is considerably less
complicated than that for matrix elements of the operators in
Eq.~(\ref{eq:quark_level_op_susy}).  Generically, the chiral expansion
for $\langle \bar{B}^{0}_{(s)} | {\cal O}_{1,aa} | B^{0}_{(s)}\rangle$
takes the form
\begin{equation}
\label{eq:O_1_chiral_generic}
 \langle \bar{B}^{0}_{(s)} | {\cal O}_{1,aa} | B^{0}_{(s)}\rangle
 \stackrel{{\rm chiral}}{\longrightarrow}
 \gamma_{1} \left ( 1 + L \right ) + {\rm analytic}\mbox{ }\mbox{ }{\rm terms},
\end{equation}
where $\gamma_{1}$ is the leading-order low-energy constant (LEC), $L$
denotes the non-analytic one-loop contributions (chiral logarithms),
and the analytic terms are from the next-to-leading-order
counter-terms in the chiral expansion.  However, for the operators in
Eq.~(\ref{eq:quark_level_op_susy}), the chiral expansion has the
generic feature:
\begin{equation}
\label{eq:O_2345_chiral_generic}
 \langle \bar{B}^{0}_{(s)} | {\cal O}_{i,aa} | B^{0}_{(s)}\rangle
 \stackrel{{\rm chiral}}{\longrightarrow}
 \gamma_{i} \left ( 1 + L \right ) + \gamma^{\prime}_{i} L^{\prime}
  + {\rm analytic}\mbox{ }\mbox{ }{\rm terms},
\end{equation}
where $i=2,3,4,5$, $\gamma_{i}$ and $\gamma^{\prime}_{i}$ are unknown
leading-order LECs, and $L$ and $L^{\prime}$ are
different one-loop chiral logarithms.  Again, the analytic terms are
from the next-to-leading-order counter-terms in the chiral expansion.
The appearance of the second non-analytic term complicates the chiral 
extrapolation in Eq.~(\ref{eq:O_2345_chiral_generic}) 
because an additional unknown parameter must be determined.
The origin of this complication is discussed in detail in
Section~\ref{sec:4q_HMChPT}.

This paper is structured as follows: in Section~\ref{sec:HMChPT} we
briefly discuss heavy meson chiral perturbation theory before turning
to the inclusion of the four-quark operators in HM$\chi$PT in
Section~\ref{sec:4q_HMChPT}. We present the results of the
next-to-leading order (NL{\cal O}) light quark mass and lattice volume
dependence of the relevant matrix elements in
Section~\ref{sec:one_loop} before concluding
(Section~\ref{sec:conclusion}). Various technical details are
relegated to the Appendices.

Whilst this work was being completed, a preprint describing similar
work appeared~\cite{Becirevic:2006me}. The conclusions of the revised
version of that work agree with those presented herein, specifically
the forms of the chiral extrapolations in
Eqs.~(\ref{eq:O_1_chiral_generic}) and
(\ref{eq:O_2345_chiral_generic}).

\section{Heavy meson chiral perturbation theory}
\label{sec:HMChPT}

The inclusion of the heavy-light mesons in chiral perturbation theory
(HM$\chi$PT) was first proposed in Refs.~\cite{Burdman:1992gh,
  Wise:1992hn,Yan:1992gz}, with the generalisation to
quenched\footnote{We do not consider the quenched theory here as
  quenched quantities are unrelated to those in QCD
  \cite{Sharpe:2001fh}.} and partially-quenched theories given in
Refs.~\cite{Booth:1995hx, Sharpe:1996qp}.  The $1/M_{P}$ ($M_{P}$ is
the mass of the heavy-light pseudo-scalar meson) and chiral
corrections were studied by Boyd and Grinstein \cite{Boyd:1995pa} in
full QCD and by Booth \cite{Booth:1994rr} in quenched QCD.  The field
appearing in this effective theory is
\begin{equation}
\label{eq:HQ_field}
 H^{(Q)}_{a} = \frac{1 + \slash\!\!\! v}{2} \left ( 
 P^{\ast (Q)}_{a,\mu} \gamma^{\mu} - P^{(Q)}_{a}\gamma_{5}\right ) ,
\end{equation}
where $P^{(Q)}_{a}$ and $P^{\ast (Q)}_{a,\mu}$ annihilate
pseudo-scalar and vector mesons containing a heavy quark $Q$ and a
light anti-quark of flavour $a$. In the heavy particle formalism, such
mesons have momentum $p^\mu=M_P v^\mu +k^\mu$ with $|k^\mu|\ll M_P$
and $v^\mu$ is the velocity of the particle.  Under a heavy quark spin
$SU(2)$ transformation $S$ and a generic light-flavour transformation
$U$ [{\it i.e.}, $U\in SU(3)$ for full QCD and $U\in SU(6|3)$ for
PQQCD (partially-quenched QCD)],
\begin{equation}
\label{eq:HQ_field_transformation}
 H^{(Q)}_{a} \longrightarrow S H^{(Q)}_{b} U^{\dagger}_{ba} .
\end{equation}
The conjugate field, which creates heavy-light mesons containing a
heavy quark $Q$ and a light anti-quark of flavour $a$, is defined as
\begin{equation}
\label{eq:barHQ_field}
 \bar{H}^{(Q)}_{a} = \gamma^{0} H^{(Q)\dagger} \gamma_{0}
 = \left ( 
 P^{\ast (Q)\dagger}_{a,\mu} \gamma^{\mu} + 
    P^{(Q)\dagger}_{a}\gamma_{5}\right )
 \frac{1 + \slash\!\!\! v}{2} ,
\end{equation}
which transforms under $S$ and $U$ as
\begin{equation}
\label{eq:barHQ_field_transformation}
 \bar{H}^{(Q)}_{a} \longrightarrow U_{ab} \bar{H}^{(Q)}_{b} S^{\dagger} .
\end{equation}
\medskip

The chiral Lagrangian for the Goldstone particles is
\begin{eqnarray}
 {\mathcal{L}}_{\mathrm{GP}} &=&
 \frac{f^{2}}{8} {\mathrm{(s)tr}} \left [ 
 \big (\partial_{\mu} \Sigma^{\dagger}\big )\big (\partial^{\mu} \Sigma \big)
 + \Sigma^{\dagger} \chi + \chi^{\dagger} \Sigma\right ]\,,
\end{eqnarray}
where $\Sigma= {\mathrm{exp}}(2 i \Phi/f)$ is the non-linear Goldstone
field, with $\Phi$ being the matrix containing the standard Goldstone
fields.  We use $f=132$~MeV.  In this work, we follow the
supersymmetric formulation of partially quenched chiral perturbation
theory [PQ$\chi$PT] \cite{Bernard:1992mk,Bernard:1994sv}.  Therefore
$\Sigma$ transforms linearly under $SU(3)_{\mathrm{L}}\otimes
SU(3)_{\mathrm{R}}$ and $SU(6|3)_{\mathrm{L}}\otimes
SU(6|3)_{\mathrm{R}}$ in full QCD and PQQCD respectively.  The symbol
``(s)tr'' in the above equation means ``trace'' in chiral perturbation
theory ($\chi$PT) and
``supertrace'' in PQ$\chi$PT where the flavour group is graded.  
The variable $\chi$ is defined as
\begin{equation}
 \chi \equiv 2 B_{0} {\mathcal{M}}_{q} = 
\frac{-2 \langle 0 |\bar{u}u+\bar{d}d|0\rangle}{f^{2}}
 {\mathcal{M}}_{q},
\end{equation}
where the quark mass matrix ${\cal M}_{q}$ is
\begin{equation}
\label{eq:full_mass_matrix}
 {\cal M}^{\mathrm{(QCD)}}_{q} = {\mathrm{diag}} 
 (m_{u},m_{u},m_{s}) ,
\end{equation}
in full QCD,
and 
\begin{equation}
\label{eq:PQ_mass_matrix}
 {\cal M}^{({\mathrm{PQQCD}})}_{q} = {\mathrm{diag}} 
 (\underbrace{m_{u},m_{u},m_{s}}_{{\mathrm{valence}}},
  \underbrace{m_j,m_j,m_r}_{{\mathrm{sea}}},
  \underbrace{m_{u},m_{u},m_{s}}_{{\mathrm{ghost}}}) ,
\end{equation}
in PQQCD.  We keep the strange quark mass different from that of the
(degenerate) up and down quarks in the valence, sea and ghost sectors.  Notice that
the flavour singlet state $\Phi_0={\mathrm{(s)tr}}(\Phi)/\sqrt{6}$ is
rendered heavy by the $U(1)_A$ anomaly in QCD and PQQCD
\cite{Sharpe:2001fh,Sharpe:2000bc} and has been integrated out.

Furthermore, the Goldstone mesons appear in the HM$\chi$PT Lagrangian
via the field
\begin{equation}
\label{eq:xi_field}
 \xi \equiv {\mathrm e}^{i\Phi/f} ,
\end{equation}
which transforms as
\begin{equation}
\label{eq:xi_field_transformation}
 \xi \longrightarrow U_{\mathrm{L}} 
  \xi U^{\dagger} = U \xi 
   U^{\dagger}_{\mathrm{R}},
\end{equation}
where $U_{\mathrm{L(R)}}$ is an element of the left-handed
(right-handed) $SU(3)$ and $SU(6|3)$ groups for QCD and PQQCD
respectively.  The HM$\chi$PT Lagrangian, to lowest order in the
chiral and $1/M_{P}$ expansion, for mesons containing a heavy quark
$Q$ and a light anti-quark of flavour $a$ is then
\begin{eqnarray}
\label{eq:HMChPT}
 {\mathcal{L}}_{\mathrm{HM\chi PT}} 
 &=& -i \, {\mathrm{tr_{D}}}
  \left (
   \bar{H}^{(Q)}_{a} v_{\mu} \partial^{\mu} H^{(Q)}_{a}
  \right )
+ \frac{i}{2}  {\mathrm{tr_{D}}}
  \left (
   \bar{H}^{(Q)}_{a} v_{\mu} \left [ \xi^{\dagger}\partial^{\mu}
   \xi + \xi
  \partial^{\mu}\xi^{\dagger}\right ]_{ab} H^{(Q)}_{b}
  \right )
 \nonumber\\
& & + \frac{i}{2} g\,
   {\mathrm{tr_{D}}}
  \left (
  \bar{H}^{(Q)}_{a} \gamma_{\mu}\gamma_{5}
   \left [ \xi^{\dagger}\partial^{\mu}\xi 
        - \xi\partial^{\mu}
  \xi^{\dagger}\right ]_{ab}   
  H^{(Q)}_{b}
  \right )
 \nonumber\\
& & + B_{\etaprime}\frac{i}{2}\gamma\,
 {\mathrm{tr_{D}}}
  \left (
  \bar{H}^{(Q)}_{a}  H^{(Q)}_{a}\gamma_{\mu}\gamma_{5}
  \right )
  {\mathrm{(s)tr}} \left [ 
  \xi^{\dagger}\partial^{\mu}\xi 
   - \xi\partial^{\mu}\xi^{\dagger}\right ] ,
\end{eqnarray}
where $B_{\etaprime}=0$ for full QCD, and $B_{\etaprime}=1$ for
PQQCD\footnote{However, since we integrate out the $\eta^{\prime}$ in
  PQQCD \cite{Sharpe:2001fh}, the coupling $\gamma$ does not appear in
  the results presented in this paper.}. The flavour (super-)trace
${\mathrm{(s)tr}}$ is taken in the appropriate flavour space and
${\mathrm{tr_{D}}}$ is the trace over Dirac space. The low energy
constant (LEC) $g$ occurring in this Lagrangian is common to both
HM$\chi$PT and partially-quenched HM$\chi$PT. Note that factors of
$\sqrt{}M_{Q}$ and $\sqrt{}M_{Q}^{\ast}$ have been absorbed into the heavy
meson fields so the $H^{(Q)}_{b}$ are of mass dimension~3/2.

The HM$\chi$PT Lagrangian for mesons containing a heavy anti-quark
$\bar{Q}$ and a light quark of flavour $a$ is obtained by applying the
charge conjugation operation to the above Lagrangian
\cite{Grinstein:1992qt}.  The field that annihilates such mesons is
\begin{equation}
 \label{eq:HQbar_field}
 H^{(\bar{Q})}_{a} = \left ( 
 P^{\ast (\bar{Q})}_{a,\mu} \gamma^{\mu} - P^{(\bar{Q})}_{a}\gamma_{5}\right ) 
 \frac{1 - \slash\!\!\! v}{2},
\end{equation}
which transforms under $S$ and $U$ as
\begin{equation}
\label{eq:HQbar_field_transformation}
 H^{(\bar{Q})}_{a} \longrightarrow U_{ab} H^{(\bar{Q})}_{b} S^{\dagger} .
\end{equation}

The effects of chiral and heavy quark symmetry breakings have been
systematically studied at next-to-leading order in
full~\cite{Boyd:1995pa} and quenched HM$\chi$PT~\cite{Booth:1994rr}.
Amongst them, the only relevant feature necessary for our calculations 
are the shifts to the masses of the heavy-light mesons.
These shifts are from the heavy quark spin breaking term
\begin{equation}
\label{eq:HQ_spin_breaking_term}
\frac{\lambda_{2}}{M_{P}}
{\mathrm{tr_{D}}} \left (
 \bar{H}^{(Q)}_{a}\sigma_{\mu\nu} H^{(Q)}_{a} \sigma^{\mu\nu}
\right ) ,
\end{equation}
and the chiral symmetry breaking terms
\begin{equation}
\label{eq:su3_violating_terms}
 \lambda_{1}B_{0}\mbox{ }{\mathrm{tr_{D}}} \left ( 
  \bar{H}^{(Q)}_{a} 
  \left [
   \xi {\mathcal{M}}_{q} \xi + 
   \xi^{\dagger} {\mathcal{M}}_{q} \xi^{\dagger}
  \right ]_{ab}
  H^{(Q)}_{b}
 \right )
+ \lambda_{1}^{\prime}B_{0}\mbox{ }{\mathrm{tr_{D}}} \left ( 
  \bar{H}^{(Q)}_{a} 
  H^{(Q)}_{a}
 \right )
 \left [
   \xi {\mathcal{M}}_{q} \xi + 
   \xi^{\dagger} {\mathcal{M}}_{q} \xi^{\dagger}
  \right ]_{bb} .
\end{equation}
We choose to use a field redefinition that allows us to work with the
effective theory in which the heavy-light pseudo-scalar mesons that
contain a heavy quark and a $u$ or $d$ valence anti-quark are
massless.  Notice that the term proportional to $\lambda^{\prime}_{1}$
in Eq.~(\ref{eq:su3_violating_terms}) causes a universal shift to all
the heavy-light meson masses.  This means that 
the propagators of the heavy mesons are as follows
\begin{equation}
\label{eq:shift_start}
 \frac{i}{2 (v\cdot k + i\epsilon)} ,\quad
 \frac{-i (g_{\mu\nu} - v_{\mu} v_{\nu})}
   {2 (v\cdot k - \Delta_{\ast} + i\epsilon)},\quad
  \frac{i}{2 (v\cdot k - \delta_{us} + i\epsilon)},\quad
{\mathrm{and}}
\quad
  \frac{-i (g_{\mu\nu} - v_{\mu} v_{\nu})}
   {2 (v\cdot k - \Delta_{\ast} - \delta_{us} + i\epsilon)} ,
\end{equation}
for $P$, $P^{\ast}$, $P_{s}$, and $P^{\ast}_{s}$, respectively.  The
mass shifts can be written in terms of the couplings in
Eqs.~(\ref{eq:HQ_spin_breaking_term}) and
(\ref{eq:su3_violating_terms}):
\begin{equation}
 \Delta_{\ast} = -8 \frac{\lambda_{2}}{M_{P}} ,
\end{equation}
and
\begin{equation}
\label{eq:delta_s_lambda}
 \delta_{us} = 2\lambda_{1} B_{0} (m_{s}-m_{u}) .
\end{equation}

In the partially quenched extension, there are two additional mass
shifts because the sea quarks masses differ from those of the valence
and ghost quarks:
\begin{equation}
\label{eq:tilde_delta_s_def}
 \delta_{jr} = M_{\tilde{P}_{s}} - M_{\tilde{P}}
      = 2\lambda_{1} B_{0} (m_r- m_j) ,
\end{equation}
and
\begin{equation}
\label{eq:delta_sea_def}
 \delta_{uj} = M_{\tilde{P}} - M_{P}
 = 2\lambda_{1} B_{0} (m_j-m_{u}) .
\end{equation}
where $\tilde{P}$ ($\tilde{P}_{s}$) is the heavy-light pseudo-scalar
meson with a $j$ ($r$) sea anti-quark.  The propagators of the heavy
mesons containing sea anti-quarks are:
\begin{equation}
  \frac{i}
   {2 (v\cdot k - \delta_{uj} + i\epsilon)},\quad
  \frac{-i (g_{\mu\nu} - v_{\mu} v_{\nu})}
   {2 (v\cdot k - \Delta_{\ast} - \delta_{uj} 
  + i\epsilon)}, \quad
  \frac{i}{2 (v\cdot k - \delta_{uj} - \delta_{jr} 
   + i\epsilon)},\quad
{\rm and}\quad
\label{eq:shift_end}
  \frac{-i (g_{\mu\nu} - v_{\mu} v_{\nu})}
   {2 (v\cdot k  - \Delta_{\ast} 
    - \delta_{uj} - \delta_{jr} + i\epsilon)}\,,
\end{equation}
for $\tilde{P}$, $\tilde{P}^*$ (vector meson with a $j$ sea
anti-quark), $\tilde{P}_s$, and $\tilde{P}^*_s$ (vector meson with an
$r$ sea anti-quark), respectively.

\section{Four-fermion operators in heavy meson chiral perturbation
  theory}
\label{sec:4q_HMChPT}

\subsection{Construction of the $\Delta B=2$ and $\Delta C=2$ operators}
\label{sec:delta-b=2-delta}

Under a chiral transformation, the four-quark operators in
Eqs.~(\ref{eq:1}) and (\ref{eq:quark_level_op_susy}) fall into two
categories:
\begin{eqnarray}
 \op_{LL} &=& \bar{h}\mbox{ }\Gamma_{LL}\mbox{ }q_{L} 
 \mbox{ }\mbox{ }\bar{h}\mbox{ }\Gamma_{LL}\mbox{ }q_{L} ,
\nonumber\\
\label{eq:two_categories}
 \op_{LR} &=& \bar{h}\mbox{ }\Gamma^{(1)}_{LR}\mbox{ }q_{L} 
 \mbox{ }\mbox{ }\bar{h}\mbox{ }\Gamma^{(2)}_{LR}\mbox{ }q_{R} ,
\end{eqnarray}
where
\begin{equation}
 q_{L,R} = \frac{1\pm\gamma_{5}}{2} q .
\end{equation}

Operators $\op_{1,aa},\,\op_{2,aa}$ and $\op_{3,aa}$ are of the
first type and transform in the symmetric $({\bf 6_L,1_R})$
representation built from the direct product $({\bf
  3_L,1_R})\otimes({\bf 3_L,1_R})=({\bf 6_L,1_R})\oplus({\bf
  \overline{3}_L,1_R})$ under chiral rotations while $\op_{4,aa}$ and
$\op_{5,aa}$ are of the second type and transform in the $({\bf
  3_L,3_R})$ representation. Here we refer to the SU(3) flavour
transformation properties, leaving the partially quenched extension to
the following subsection. Note that the colour indices in
Eq.~(\ref{eq:quark_level_op_susy}) are relevant to short-distance
physics, and hence play no role in the chiral properties of these
operators \cite{Becirevic:2004qd}.  Treating $\Gamma_{LL}$,
$\Gamma^{(1)}_{LR}$ and $\Gamma^{(2)}_{LR}$ as spurions transforming
as
\begin{eqnarray}
 \Gamma_{LL} \longrightarrow S\mbox{ }\Gamma_{LL}\mbox{ }U^{\dagger}_{L} ,
\nonumber\\
 \Gamma^{(1)}_{LR} \longrightarrow S\mbox{ }
    \Gamma^{(1)}_{LR}\mbox{ }U^{\dagger}_{L} ,
\nonumber\\
 \Gamma^{(2)}_{LR} \longrightarrow S\mbox{ }
   \Gamma^{(2)}_{LR}\mbox{ }U^{\dagger}_{R} ,
\end{eqnarray}
the operators in Eq.~(\ref{eq:two_categories}) remain invariant under
heavy-quark spin and chiral rotations.  We then find that
the bosonisation of the operators in Eqs.~(\ref{eq:1}) and (\ref{eq:quark_level_op_susy})
is given by
\begin{eqnarray}
  \label{eq:4}
  \op_{i,aa}^{\rm HM\chi PT} &=&
  \sum_x
  \Bigg\{
\alpha_{i,x}^{(1)}\TrD \left [ \left ( \xi \bar{H}^{(h)} \right )_{a} 
                         \Gamma\, \Xi_x\right ]
             \TrD \left [ \left ( \xi H^{(\bar{h})} \right )_{a} 
                        \Gamma\, \Xi_x^\prime \right ]
+\alpha_{i,x}^{(3)}\TrD \left [ \left ( \xi \bar{H}^{(h)} \right )_{a} 
                         \Gamma\, \Xi_x \left ( \xi H^{(\bar{h})} \right )_{a} 
                        \Gamma\, \Xi_x^\prime \right ]
\Bigg\}\,,
\end{eqnarray}
for $i=1,2,3$ where $\Gamma=\Gamma_1=\Gamma_2$ in Eq.~(\ref{eq:2}) and
$\Xi_x$ and $\Xi_x^\prime$ are all possible pairs of Dirac
structures\footnote{An overcomplete list of the possible pairs of
  structures is: $\{\Xi_x,\Xi_x^\prime\}=\{\{1,1\}, \{1,\vslash \},
  \{\vslash,\vslash \}, \{\gamma_\mu,\gamma^\mu\}, \{\gamma_\mu
  \vslash,\gamma^\mu\}, \{\gamma_\mu\vslash,\gamma^\mu\vslash\},
  \{\sigma_{\mu\nu},\sigma^{\mu\nu}\}$,
  $\{\sigma_{\mu\nu}\vslash,\sigma^{\mu\nu}\},
  \{\sigma_{\mu\nu}\vslash,\sigma^{\mu\nu}\vslash\}\}$, their
  permutations, and possible combinations with $\gamma_5$.  There is
  some redundancy here as the equations of motion of the heavy meson
  fields, $\vslash H_a^{(Q)}=H_a^{(Q)}$ etc., relate various terms in
  Eqs.~(\ref{eq:4}) and~(\ref{eq:HMXPTops45}).}.  For $i=4,5$ the
HM$\chi$PT operators are
\begin{eqnarray}
  \label{eq:HMXPTops45}
  \op_{i,aa}^{\rm HM\chi PT} &=&
  \sum_x
  \Bigg\{
\alpha_{i,x}^{(1)}\TrD \left [ \left ( \xi \bar{H}^{(h)} \right )_{a} 
                         \Gamma_1\, \Xi_x\right ]
             \TrD \left [ \left ( \xi^\dagger H^{(\bar{h})} \right )_{a} 
                        \Gamma_2\, \Xi_x^\prime \right ]
+\alpha_{i,x}^{(2)}\TrD \left [ \left ( \xi \bar{H}^{(h)} \right )_{a} 
                         \Gamma_2\, \Xi_x\right ]
             \TrD \left [ \left ( \xi^\dagger H^{(\bar{h})} \right )_{a} 
                        \Gamma_1\, \Xi_x^\prime \right ]
\nonumber
\\
 & & 
%\hspace*{1cm}
+\alpha_{i,x}^{(3)}\TrD \left [ \left ( \xi \bar{H}^{(h)} \right )_{a} 
 \Gamma_1\, \Xi_x \left ( \xi^\dagger H^{(\bar{h})} \right )_{a} 
                        \Gamma_2\, \Xi_x^\prime \right ]
+\alpha_{i,x}^{(4)}\TrD \left [ \left ( \xi \bar{H}^{(h)} \right )_{a} 
 \Gamma_2\, \Xi_x  \left ( \xi^\dagger H^{(\bar{h})} \right )_{a} 
                        \Gamma_1\, \Xi_x^\prime \right ]
\nonumber
\\
 & & 
%\hspace*{1cm}
+\alpha_{i,x}^{(5)}\TrD \left [ \left ( \xi^\dagger \bar{H}^{(h)} \right )_{a} 
                         \Gamma_1\, \Xi_x\right ]
             \TrD \left [ \left ( \xi H^{(\bar{h})} \right )_{a} 
                        \Gamma_2\, \Xi_x^\prime \right ]
+\alpha_{i,x}^{(6)}\TrD \left [ \left ( \xi^\dagger \bar{H}^{(h)} \right )_{a} 
                         \Gamma_2\, \Xi_x\right ]
             \TrD \left [ \left ( \xi H^{(\bar{h})} \right )_{a} 
                        \Gamma_1\, \Xi_x^\prime \right ]
\nonumber
\\
 & & 
%\hspace*{1cm}
+\alpha_{i,x}^{(7)}\TrD \left [ \left ( \xi^\dagger \bar{H}^{(h)} \right )_{a} 
     \Gamma_1\, \Xi_x \left ( \xi H^{(\bar{h})} \right )_{a} 
                        \Gamma_2\, \Xi_x^\prime \right ]
+\alpha_{i,x}^{(8)}\TrD \left [ \left ( \xi^\dagger \bar{H}^{(h)} \right )_{a} 
       \Gamma_2\, \Xi_x  \left ( \xi H^{(\bar{h})} \right )_{a} 
                        \Gamma_1\, \Xi_x^\prime \right ]
\Bigg\}\,.
\end{eqnarray}
The positions in the above operators in which the arbitrary Dirac
structures, $\Xi_x$ and $\Xi_x^\prime$, are inserted is constrained by
the heavy-quark spin symmetry~\cite{Isgur:1989ed, Isgur:1989vq}. 
Notice that in general both single and
double Dirac trace operators must be considered.

Performing the Dirac traces for the particular $\Gamma$,
$\Gamma_{1,2}$ in Eqs.~(\ref{eq:1}) and
(\ref{eq:quark_level_op_susy}), and keeping only the terms that will
contribute to the matrix elements we consider, leads to the following
set of operators involving the individual heavy meson fields:
\begin{eqnarray}
 \op^{\rm HM\chi PT}_{1,aa} &=&  \beta_{1} \left [ 
            \left ( \xi P^{(h)\dagger} \right )_{a}
            \left ( \xi P^{(\bar{h})} \right )_{a}
          + \left ( \xi P^{\ast (h)\dagger}_{\mu} \right )_{a}
            \left ( \xi P^{\ast (\bar{h}),\mu} \right )_{a}
                 \right ] \,,
\nonumber\\
\label{eq:tree_level_ops}
 \op^{\rm HM\chi PT}_{2(3),aa} &=&  \beta_{2(3)} 
            \left ( \xi P^{(h)\dagger} \right )_{a}
            \left ( \xi P^{(\bar{h})} \right )_{a} 
            +  \beta_{2(3)}^\prime
            \left ( \xi P^{\ast (h)\dagger}_{\mu} \right )_{a}
            \left ( \xi P^{\ast (\bar{h}),\mu} \right )_{a}\,,
\\
 \op^{\rm HM\chi PT}_{4(5),aa} &=&  \beta_{4(5)}  
            \left ( \xi P^{(h)\dagger} \right )_{a}
            \left ( \xi^{\dagger} P^{(\bar{h})} \right )_{a} 
           + \hat\beta_{4(5)}
            \left ( \xi^{\dagger} P^{(h)\dagger} \right )_{a}
            \left ( \xi P^{(\bar{h})} \right )_{a} 
\nonumber\\
&&           + \beta_{4(5)}^\prime
            \left ( \xi P^{\ast (h)\dagger}_{\mu} \right )_{a}
            \left ( \xi^{\dagger} P^{\ast (\bar{h}),\mu} \right )_{a}
           + \hat\beta_{4(5)}^\prime
            \left ( \xi^{\dagger} P^{\ast (h)\dagger}_{\mu} \right )_{a}
            \left ( \xi P^{\ast (\bar{h}),\mu} \right )_{a}\,,
\nonumber
\end{eqnarray}
where the $\beta_i,\, \beta_i^\prime,\hat\beta_i$, and
$\hat\beta_i^\prime$ are linear combinations of the various
$\alpha_{i,x}^{(j)}$ appearing in Eqs. (\ref{eq:4}) and
(\ref{eq:HMXPTops45}).

It is important to note that in the above equation, the operator
${\cal O}^{\rm HM\chi PT}_{1,aa}$ behaves somewhat differently from
the other operators as only a single LEC, $\beta_{1}$, occurs.  This
greatly simplifies any chiral extrapolation of corresponding lattice
data for neutral heavy-light meson mixing in the Standard Model, as
confirmed by the one-loop results presented in the next section. We
stress that this simplification is not obvious from the operator
structure in Eq.~(\ref{eq:4}) and is particular to the $V-A$ structure
of the Standard Model currents.  In general, one would expect
from Eqs.~(\ref{eq:4}) and (\ref{eq:HMXPTops45}) that operators for
pseudo-scalar and vector meson mixing processes are accompanied by
different LECs.  This is the case for all the
non-Standard-Model operators, as shown in
Eq.~(\ref{eq:tree_level_ops}).

To understand the origin of the above simplification in the Standard
Model operator $\op^{\rm HM\chi PT}_{1,aa}$, we turn to heavy quark
effective theory (HQET) 
\cite{Grinstein:1990mj, Eichten:1989zv, Georgi:1990um}.  
In this effective
theory, the operators that produce the same matrix elements as those
in Eq.(\ref{eq:2}) are \cite{Flynn:1990qz}
\begin{eqnarray}
  \label{eq:5}
  \op_{i,aa}^{\rm HQET} &=& \tilde{Q}\Gamma_1 q_a\, Q^\dagger \Gamma_2 q_a
  + Q^\dagger \Gamma_1 q_a \, \tilde{Q}\Gamma_2 q_a\,, 
\end{eqnarray}
where $\Gamma_{1,2}$ are the appropriate Dirac and colour structures
from Eq.~(\ref{eq:quark_level_op_susy}). Here, $Q$ and $\tilde{Q}$
denote fields annihilating a heavy quark and heavy anti-quark,
respectively (these fields do not create the corresponding
anti-particles).  Additional terms in HQET which create two heavy
quarks or annihilate two heavy anti-quarks will not contribute to
neutral heavy-meson mixing and are ignored.

The standard model operator in HQET, $\op_{1,aa}^{\rm HQET}$,
satisfies the relation
\begin{eqnarray}
  \label{eq:6}
  \left\{S_Q^3, \op_{1,aa}^{\rm HQET} \right\} | P \rangle 
= \left\{S_Q^3, \op_{1,aa}^{\rm HQET} \right\} | \bar{P} \rangle
&=& 0\,,
\end{eqnarray}
where $|P\rangle$ is pseudo-scalar heavy-light meson state, and
$|\bar{P}\rangle$ is the state of its anti-particle.  The operator
\begin{equation}
\label{eq:HQ_spin_op}
 S_Q^3=\epsilon^{ij3}[Q^\dagger \sigma_{ij}Q - \tilde{Q}
\sigma_{ij} \tilde{Q}^\dagger] ,
\end{equation}
is the heavy quark spin operator \cite{Savage:1990di} that changes the
spin of the heavy-light meson state by one. Therefore,
Eq.~(\ref{eq:6}) implies that the mixing matrix elements for vector
and pseudo-scalar heavy-light mesons are equal and opposite
\cite{Grinstein:1992qt} in the heavy-quark limit.  This symmetry is
reflected in HM$\chi$PT, leading to the result for $\op^{\rm HM\chi
  PT}_{1,aa}$ in Eq.~(\ref{eq:tree_level_ops}).

For the non-Standard-Model operators, it is straightforward to show
that
\begin{equation}
\label{eq:non_sm_ops_anti_comm}
  \left\{S_Q^3, \op_{i,aa}^{\rm HQET} \right\} | P \rangle \not= 0 , \, \,
  \left\{S_Q^3, \op_{i,aa}^{\rm HQET} \right\} | \bar{P} \rangle \not= 0 ,
\end{equation}
and
\begin{equation}
\label{eq:non_sm_ops_comm}
  \left [ S_Q^3, \op_{i,aa}^{\rm HQET} \right ] | P \rangle \not= 0 , \, \,
  \left [ S_Q^3, \op_{i,aa}^{\rm HQET} \right ] | \bar{P} \rangle \not= 0 ,
\end{equation}
where $i=2,3,4,5$. This means that the pseudo-scalar and vector meson
mixing processes via these operators are not proportional to each
other, hence the appearance of the terms accompanied by
$\beta^{\prime}_{2,3,4,5}$ and $\hat{\beta}^{\prime}_{4,5}$ in
Eq.~(\ref{eq:tree_level_ops}).

We end this subsection by noting that
equations of motion for the heavy quark~\cite{Becirevic:2001xt} result in
$\op_{3,aa} =-\op_{1,aa}/2-\op_{2,aa}$, and can be used to relate
some of the LECs in Eq.~(\ref{eq:tree_level_ops}).

\subsection{Partially-quenched extensions}
\label{sec:part-quench-extens}

In the partially-quenched theory, the operator matching is analogous 
because the QCD operators considered on the lattice involve
only valence quarks.  Since HM$\chi$PT is contained within
partially-quenched HM$\chi$PT, the LECs occurring in the four-quark
operators of both theories are the same for the quantities we
consider.

In the partially quenched case, the $\Delta B=2$ and $\Delta C=2$
operators transform in the symmetric tensor product of two fundamental
representations of SU(6$|$3), a {\bf 42}-dimensional representation.
The operators arising from QCD can be simply embedded in this larger
representation with no mixing into different representations. 
In most cases (and herein) it is sensible to choose the quark
``charges'' such that the operators are purely valence, but any other
element of this representation suffices.

\section{Neutral meson mixing matrix elements}
\label{sec:one_loop}

\begin{figure}[!t]
  \centering
\includegraphics[width=0.35\columnwidth]{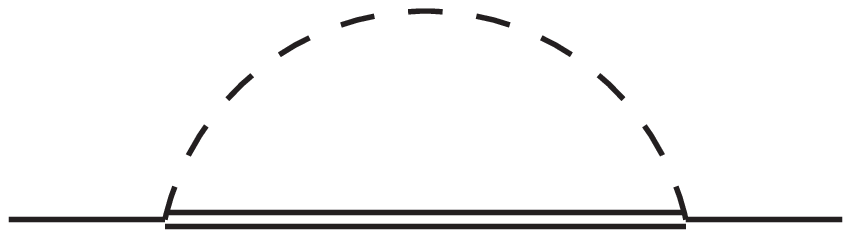}
\hspace*{1mm}
\includegraphics[width=0.225\columnwidth]{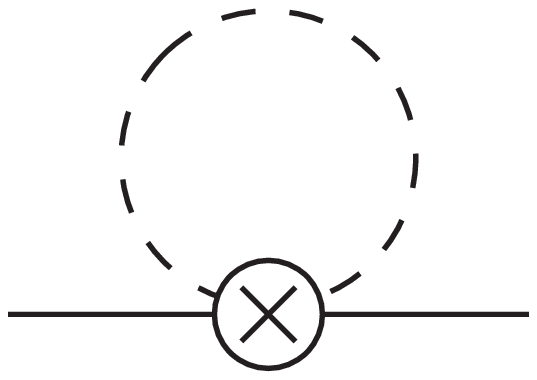}
\hspace*{1mm}
\includegraphics[width=0.35\columnwidth]{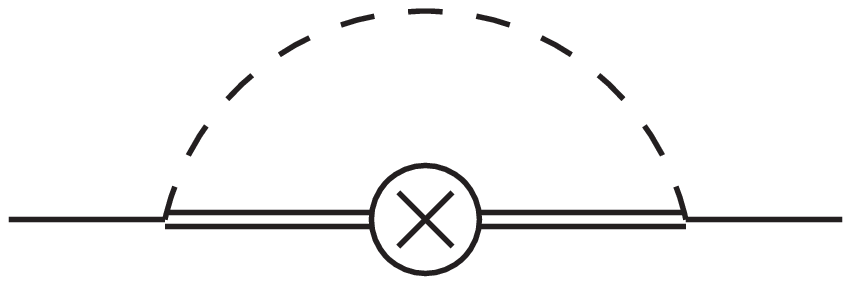} 
\\
(a)\hspace*{5.1cm}(b)\hspace*{5.1cm}(c)
  \caption{Diagrams contributing to the matrix elements of four-quark
    operators at NLO in the chiral expansion. Solid, double and dashed
    lines correspond to propagators of pseudo-scalar and vector
    heavy-light mesons, and Goldstone mesons, respectively. The
    crossed circle denotes the four-quark operator and diagram (a) is
    the wave-function renormalisation.}
  \label{fig:oneloop}
\end{figure}

Calculations at NLO in the chiral expansion require the evaluation of
the one-loop diagrams shown in Figure~\ref{fig:oneloop}. We perform
these calculations both at infinite volume and in a cubic spatial box
of dimensions $L^3$ (the time extent is assumed to be infinite). In
this section we summarise the results, relegating details of the
calculations to Appendix~\ref{sec:integrals-sums}.  

For the standard model operator we find the following matrix elements
\begin{eqnarray}
  \langle \bar{B}^0 | \op_{1,dd}|B^0\rangle &=&
  \beta_1\left(1+ {\cal T}^{(1)}_{d} + \frac{{\cal 
        W}_{\bar{B}^0}+{\cal W}_{B^0}}{2}+{\cal
      Q}^{(1)}_{d} \right) + {\rm analytic \, \, terms}\,, 
\nonumber
\\
  \label{eq:7}
\\
  \langle \bar{B}^0_s | \op_{1,ss}|B^0_s\rangle &=&
  \beta_1\left(1+ {\cal T}^{(1)}_{s} + \frac{{\cal
        W}_{\bar{B}^0_s}+{\cal W}_{B^0_s}}{2}+{\cal
      Q}^{(1)}_{s} \right) + {\rm analytic \, \, terms}\, .
\nonumber
\end{eqnarray}
The wave-function contributions, ${\cal W}_M$, and
tadpole- and sunset-type operator renormalisations, ${\cal
  T}^{(i)}_{a}$ and ${\cal Q}^{(i)}_{a}$ (diagrams (b) and (c) in
Figure~\ref{fig:oneloop}, respectively) are non-analytic functions of
the light quark mass and lattice volume and are defined in
Appendix~\ref{sec:loop-contributions}.
The ``analytic terms'' here include Goldstone meson mass squared
terms, a term $\sim \alpha_s(M_b)/4\pi$ (arising from mixing) and a
term $\sim \Lambda_{\rm QCD}/M_b$. The $\sim \Lambda_{\rm QCD}/M_b$
term arises from $1/M_b$ terms in the Lagrangian (those from Eq.
(\ref{eq:HQ_spin_breaking_term}) are included) and from additional
$1/M_b$-suppressed HM$\chi$PT operators that match onto the QCD
operators. We note that at higher orders, the simple relation between
the $B$ and $B^\ast$ matrix elements of the Standard Model operator
will break down. 
Although we
present results specifically in the $B$-meson systems, note that they
are also applicable to $D$-meson systems, under the assumption that
the charm-quark mass is large enough compared to $\Lambda_{\rm QCD}$.

Parameterising the $\Delta B=2$ matrix elements in the standard form
\cite{Gabbiani:1996hi} ($f_{B_q}$ is the weak decay constant defined
through $\langle 0|\bar{b}\gamma_\mu\gamma_5 q|B_q(\vec{p})\rangle=i
p_\mu f_{B_q}$),
\begin{eqnarray}
  \label{eq:8}
  \langle \bar{B}^0_q | \op_{1,qq}|B^0_q\rangle &=&
  \frac{8}{3}M_{B_q}^2 f_{B_q}^2 B^{(1)}_{B_q}(\mu)\,,
\\
\label{eq:9}
  \langle \bar{B}^0_q | \op_{i,qq}|B^0_q\rangle &=&
  \eta_i R^2 M_{B_q}^2
  f_{B_q}^2 B^{(i)}_{B_q}(\mu)\quad {\rm for}\,\, i=2,\ldots,5\,,
\end{eqnarray}
($R=\frac{M_{B_q}}{m_b(\mu)+m_q(\mu)}$ and $\eta_2=-\frac{5}{3}$,
$\eta_3=\frac{1}{3}$, $\eta_4=2$, $\eta_5=\frac{2}{3}$) the bag
parameters, $B^{(1)}_{B_q}(\mu)$ agree with those derived in
Refs.~\cite{Grinstein:1992qt,Arndt:2004bg} where the relevant
expressions for $f_{B_q}$ are also provided.

For the additional operators that contribute to the $B$-meson mixing
processes beyond the Standard Model, we obtain:
\begin{eqnarray}
  \langle \bar{B}^0 | \op_{2(3),dd}|B^0\rangle &=&
  \beta_{2(3)}\left(1+ {\cal T}^{(2(3))}_{d} + \frac{{\cal 
        W}_{\bar{B}^0}+{\cal W}_{B^0}}{2} \right) + 
\beta_{2(3)}^\prime {\cal Q}^{(2(3))}_{d} +{\rm analytic \, \, terms}\,, 
\nonumber
\\
\nonumber
\\
  \langle \bar{B}^0_s | \op_{2(3),ss}|B^0_s\rangle &=&
  \beta_{2(3)}\left(1+ {\cal T}^{(2(3))}_{s} + \frac{{\cal 
        W}_{\bar{B}^0_s}+{\cal W}_{B^0_s}}{2} \right) + 
\beta_{2(3)}^\prime {\cal Q}^{(2(3))}_{s} +{\rm analytic \, \, terms}\,. 
\nonumber 
\\
 \label{eq:10}
\\
  \langle \bar{B}^0 | \op_{4(5),dd}|B^0\rangle &=&
  \left[\beta_{4(5)}+\hat\beta_{4(5)}\right] 
  \left(1+ {\cal T}^{(4(5))}_{d} + \frac{{\cal 
        W}_{\bar{B}^0}+{\cal W}_{B^0}}{2} \right) + 
\left[\beta_{4(5)}^\prime+\hat\beta_{4(5)}^\prime\right] 
 {\cal Q}^{(4(5))}_{d} +{\rm analytic \, \, terms}\,, 
\nonumber
\\
\nonumber 
\\
  \langle \bar{B}^0_s | \op_{4(5),ss}|B^0_s\rangle &=&
  \left[\beta_{4(5)}+\hat\beta_{4(5)}\right] 
\left(1+ {\cal T}^{(4(5))}_{s} + \frac{{\cal 
        W}_{\bar{B}^0_s}+{\cal W}_{B^0_s}}{2} \right) + 
\left[\beta_{4(5)}^\prime+\hat\beta_{4(5)}^\prime\right]
 {\cal Q}^{(4(5))}_{s} +{\rm analytic \, \, terms}\,. 
\nonumber
\end{eqnarray}
The terms $\sim{\cal Q}_q^{(i)}$ arising from the sunset diagrams
[Fig.~\ref{fig:oneloop}(c)] involve the neutral heavy-light vector
meson mixing amplitudes.  As discussed in the preceding section, it is
only in the case of ${\cal O}_{1,qq}$ that these amplitudes are
related to those of the pseudo-scalar heavy-light mesons. For
$i=2,3,4,5$ these terms are consequently accompanied by different
LECs. The analytic terms in the above expressions
depend on the renormalisation scale is such a way as to cancel the
scale dependence of the non-analytic loop contributions.

The matrix elements in Eqs.~(\ref{eq:7}) and (\ref{eq:10}) also
determine the $B_s$ decay width differences [Eq~(\ref{eq:3})]. The
extrapolation in light quark mass and lattice volume is more involved
here than for the matrix elements determining Standard Model
oscillations and have not been accounted for in the existing
unquenched lattice calculations \cite{Gimenez:2000jj,
  Hashimoto:2001zq}. Direct calculations of the ratio of the matrix
element of $\op_{1,ss}$ to that of $\op_{2,ss}$ do not help in this
regard as the non-analytic behaviour does not simplify.

\begin{figure}[!th]
  \centering
  \includegraphics[width=0.46\columnwidth]{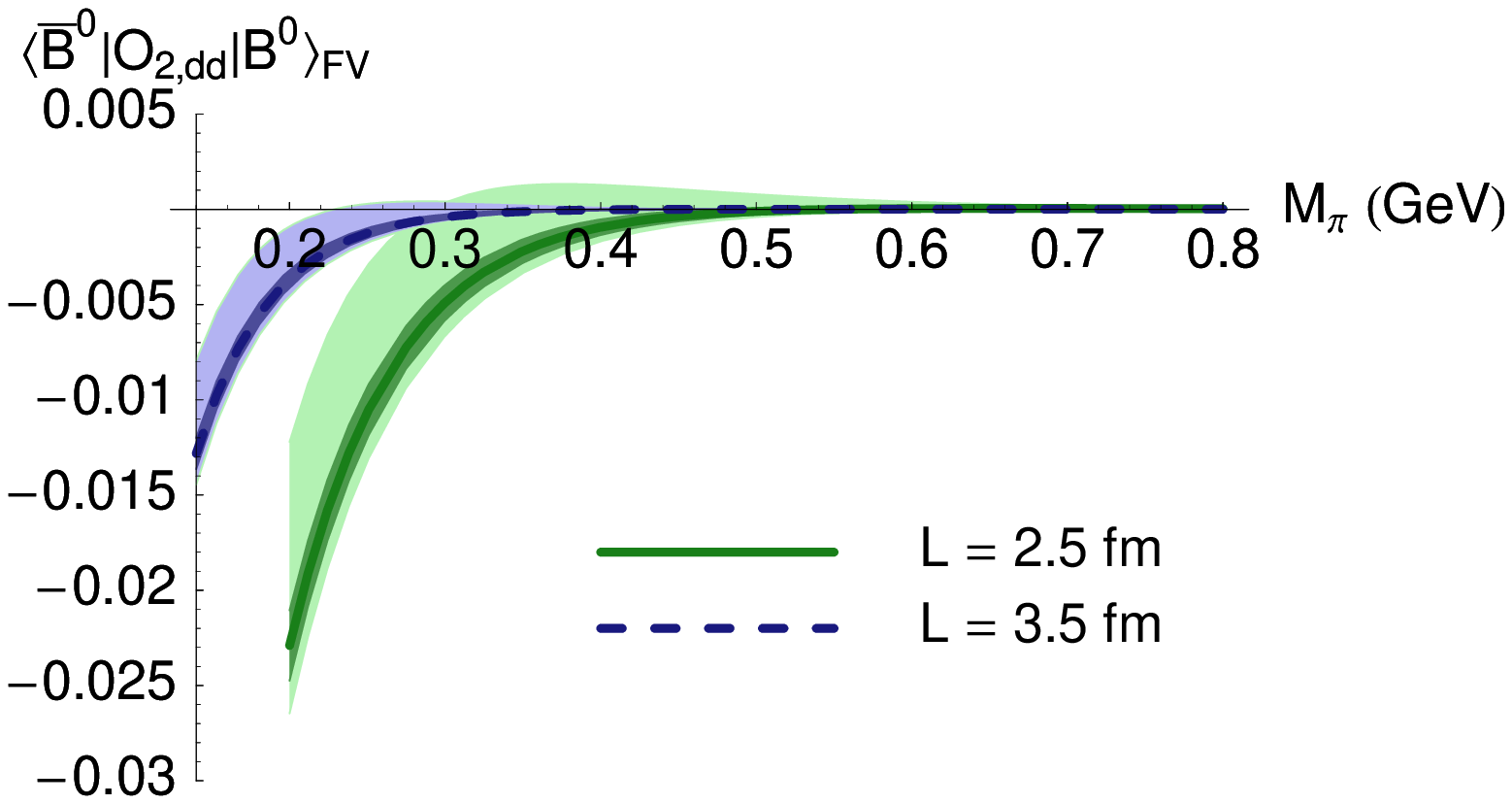}
  \includegraphics[width=0.46\columnwidth]{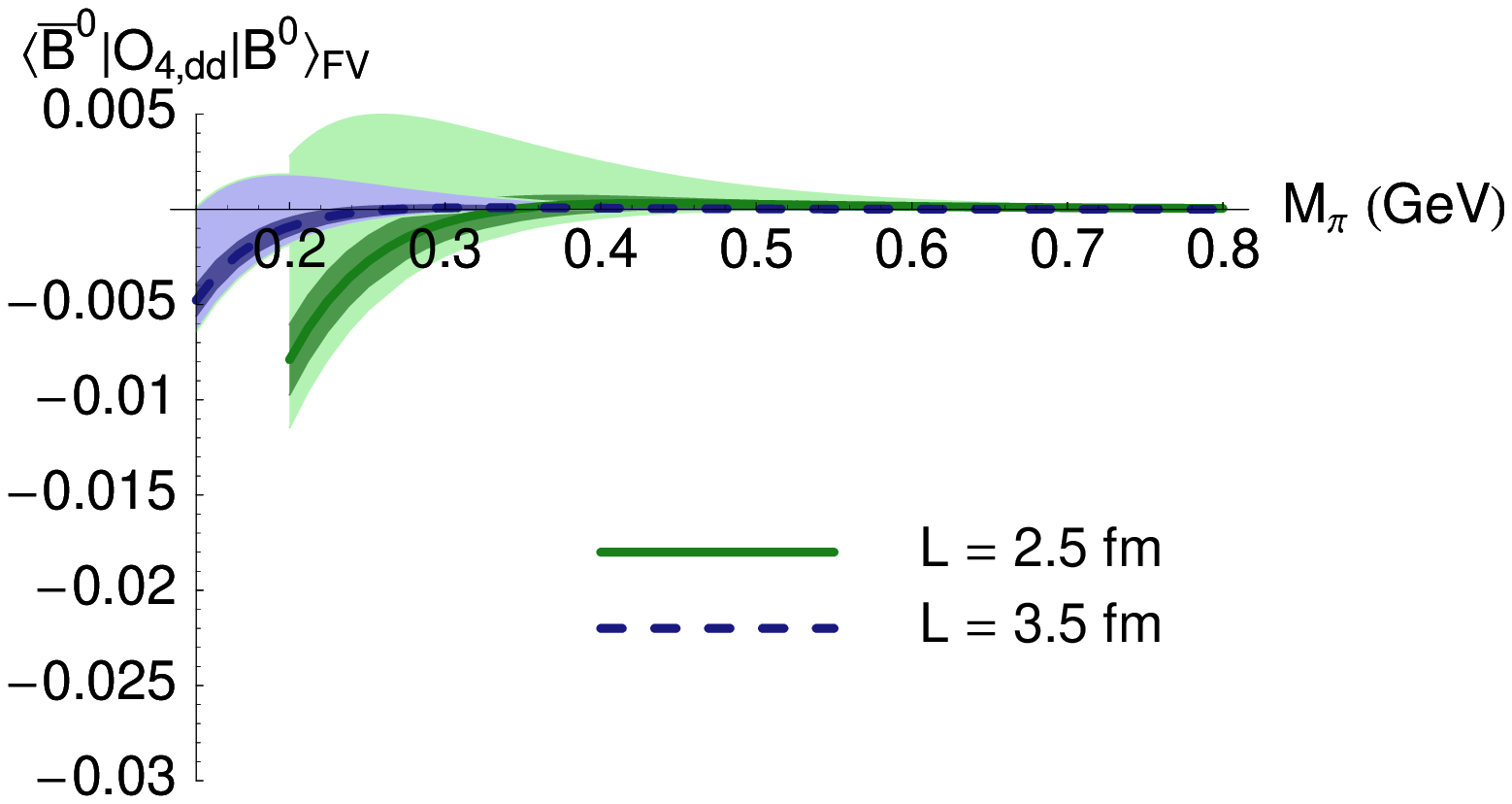}
  \caption{Finite volume effects in mixing matrix elements of the
    operators ${\cal O}_{i,dd}$ for $i=2,4$ for two lattice volumes,
    $L=2.5,\,3.5$~fm. The central curve correpsonds to $g=0.4$ and
    $\beta_i^\prime/\beta_i=1$ while the inner (darker) and outer
    (lighter) shaded regions correspond to variation of $0.3<g<0.5$
    and $|\beta_i^\prime/\beta_i|<2$. The curves terminate at
    $m_\pi\,L=2.5$ where $p$-regime chiral perturbation theory becomes
    unreliable.}
  \label{fig:FVd}
\end{figure}

\begin{figure}[!th]
  \centering
  \includegraphics[width=0.46\columnwidth]{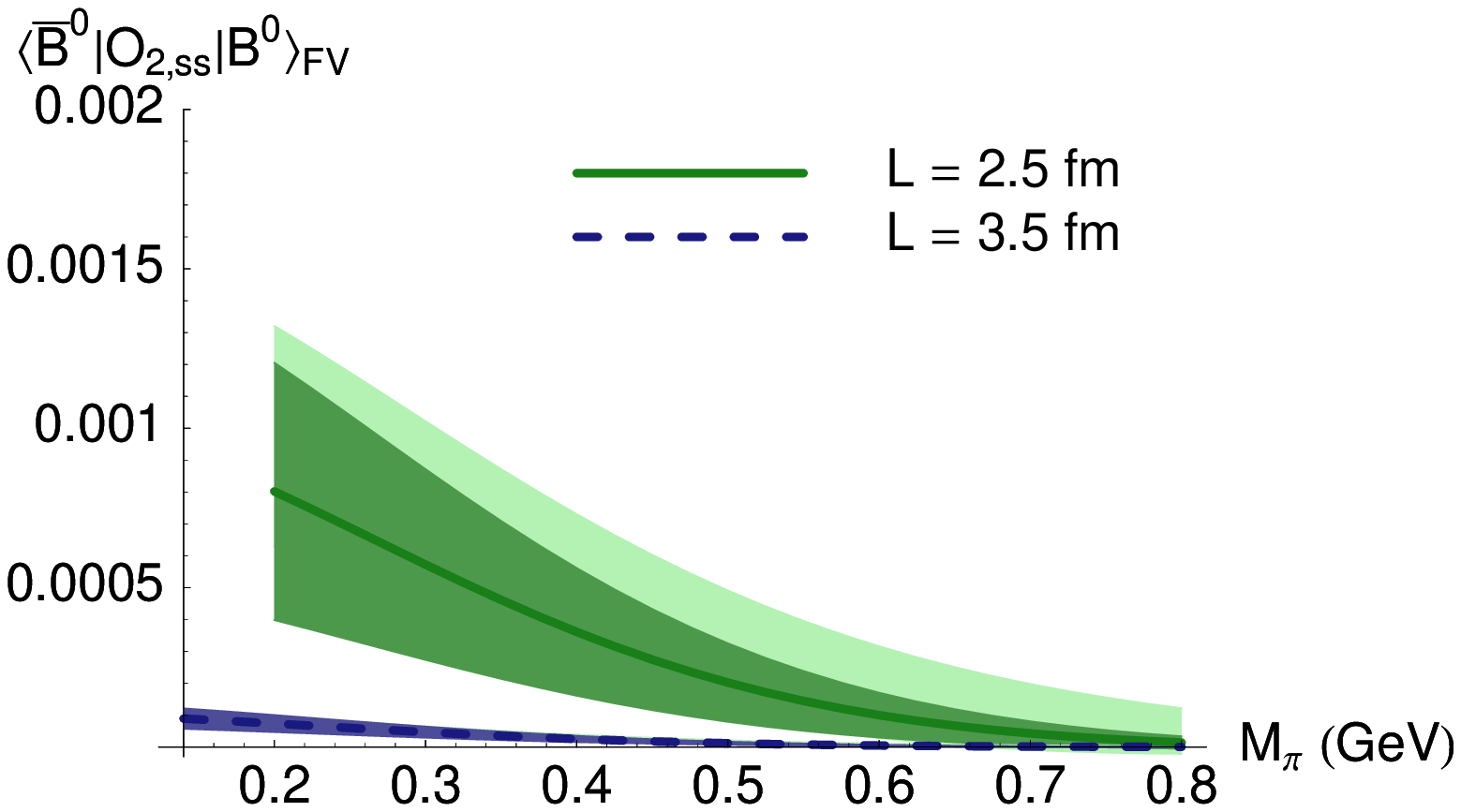}
  \includegraphics[width=0.46\columnwidth]{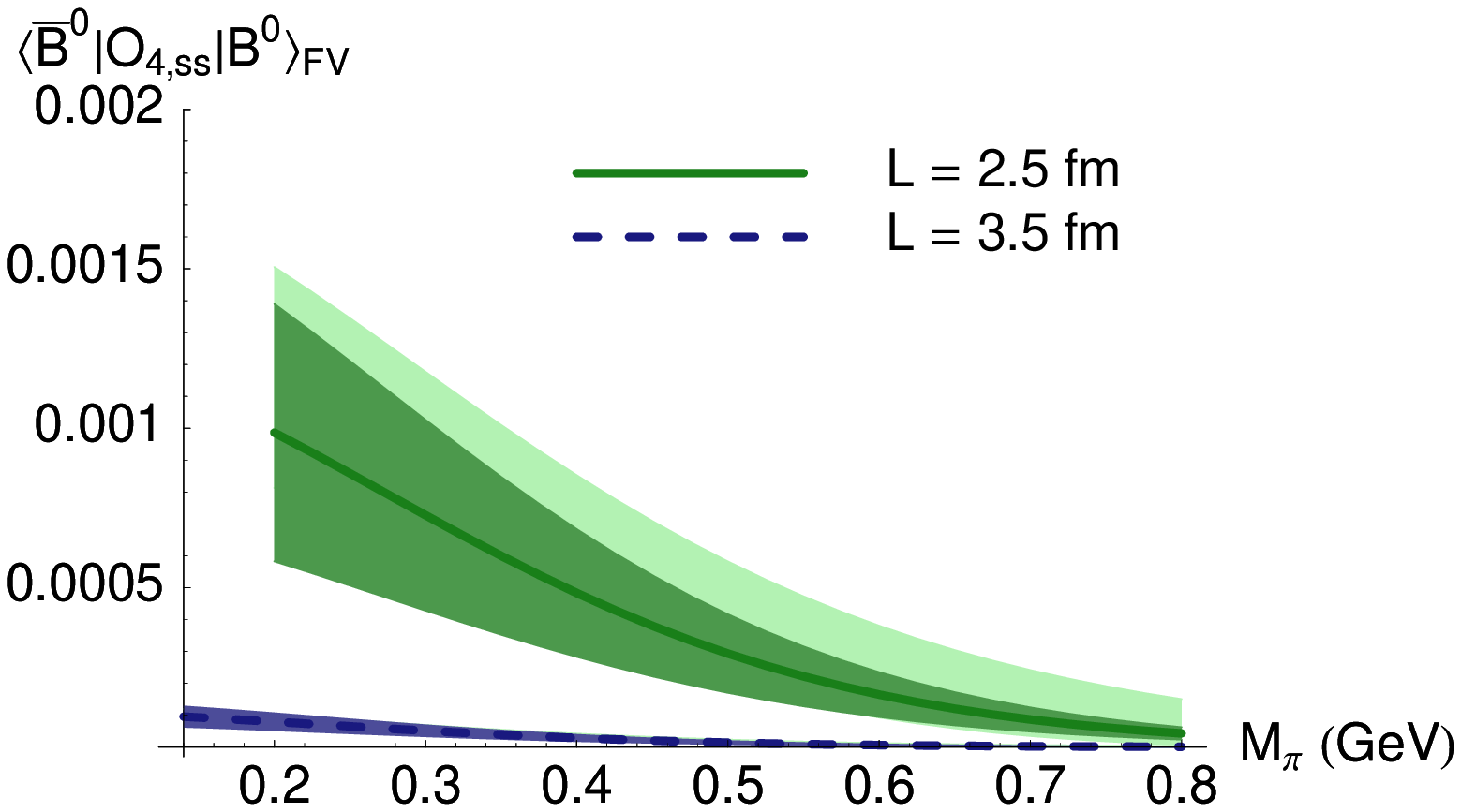}
  \caption{Finite volume effects in mixing matrix elements of the
    operators ${\cal O}_{i,ss}$ for $i=2,4$ for two lattice volumes,
    $L=2.5,\,3.5$~fm. Details are as in Fig.~\protect{\ref{fig:FVd}}.}
  \label{fig:FVs}
\end{figure}

At present, we can only study the finite volume and light quark mass
effects described in these formulae cursorily as there is very little
lattice data to use for such a task in any reliable manner. This will
hopefully change in the near future; the recent calculations of
\cite{Dalgic:2006gp} are encouraging (we note, however, that our
results imply that more than three light quark masses are needed for
the NLO chiral extrapolation of matrix elements of ${\cal O}_{i,aa}$
for $i=2,3,4,5$).  As a guide to the importance of such effects, we
present representative results for the ${\cal O}_{2,4}$ matrix
elements. Results for the Standard Model operators have been discussed
in Ref.~\cite{Arndt:2004bg}. Taking the QCD limit for definiteness,
Figures \ref{fig:FVd} and \ref{fig:FVs} explore the dependence of the
finite volume shifts in the matrix elements of $\op_{2,dd}$,
$\op_{2,ss}$ $\op_{4,dd}$ and $\op_{4,ss}$, normalised by their
tree-level values, on the pion mass for two different volumes,
$L=2.5,\,3.5$~fm. In each figure we plot the ratio
\begin{equation}
  \label{eq:23}
  \langle \overline{B}^0_{(s)}|{\cal O}_{i,ff}|B^0_{(s)}\rangle_{\rm
    FV}
= \frac{ \langle \overline{B}^0_{(s)}|{\cal
    O}_{i,ff}|B^0_{(s)}\rangle(L)- \langle \overline{B}^0_{(s)}|{\cal
    O}_{i,ff}|B^0_{(s)}\rangle(\infty)}{\langle
  \overline{B}^0_{(s)}|{\cal O}_{i,ff}|B^0_{(s)}\rangle_{\rm tree}}
\,.
\end{equation}
We fix $f=0.132$~GeV and $\Delta_\ast=50$~MeV (variation with
$\Delta_\ast$ is small and similar to that found for the Standard
Model operator \cite{Arndt:2004bg}, with the FV effect decreasing with
increasing $\Delta_\ast$). Using recent CLEO measurements
\cite{Ahmed:2001xc,Anastassov:2001cw}, the coupling $g$ is taken to be
$0.3<g^2<0.5$ with a central value of $g^2=0.4$; variation in $g$ is
indicated in the figures by the inner (darker) shaded regions.  For
each quantity, we vary the ratio of $B^{\ast}$ to $B$ LECs over a
reasonable range, taking $|\beta_2^\prime/\beta_2|<2$ and
$|(\beta_4^\prime+\hat\beta_4^\prime)/(\beta_4+\hat\beta_4)|<2$
(naturalness would suggest these ratios should be of order unity, but
for simplicity we also allow smaller values). In each figure, the
central curve corresponds to a ratio of unity and the outer (lighter)
shaded region to this variation. As can be seen, effects of the finite
volume on ${\cal O}_{i,dd}$ are similar in size to those found for the
Standard Model operator in Ref.~\cite{Arndt:2004bg}. Effects for the
strange operators are considerably suppressed as pion loops do not
contribute to these matrix elements in the QCD limit.

Before concluding, it is useful to consider how lattice spacing
artefacts will enter the above expressions. Since calculations here
involve both light and heavy quarks, there are two types of
discretisation effects. The effects of light quark discretisation are
very simple to incorporate; at a particular lattice spacing, $a$, the
masses of the various Goldstone mesons in chiral loops are shifted
from their continuum values (and these shifted masses should be used
in in fits to lattice data using the above partially-quenched
expressions) and the various counter-terms become polynomials in the
lattice spacing.  In general this polynomial will contain all powers
of $a$, but if both the light-quark action and the four quark operator
are improved (or a discretisation satisfying the Ginsparg-Wilson
relation \cite{Ginsparg:1981bj} is used for the valence quarks
\cite{Chen:2006wf}), the leading corrections in $a$ can be eliminated.
For most foreseeable calculations, this is the extent of
discretisation effects at the order to which we have worked (we assume
that $a\,\Lambda_{\rm QCD} \alt m_q/\Lambda_{\rm QCD}$).  However, if
a heavy quark action is used that breaks heavy quark spin symmetry at
${\cal O}(a)$, additional complications will arise as this symmetry
can no longer be used to constrain the form of the EFT operator,
${\cal O}^{\rm HM\chi PT}_{1,aa}$; $B_{(s)}$--$\bar{B}_{(s)}$ and
$B^\ast_{(s)}$--$\bar{B}^\ast_{(s)}$ matrix elements are no longer
related.  The resulting mass and volume dependence of matrix elements
of this operator will become more complicated, resembling instead that
of the matrix elements of the other operators in Eqs.~(\ref{eq:4}) and
(\ref{eq:HMXPTops45}).\footnote{Additional complications beyond the
  scope of this work may arise if the Kogut-Susskind fermion action is
  used for the light quarks.}

\section{Conclusion}
\label{sec:conclusion}

We have considered the matrix elements of four-quark operators
relevant for heavy-light neutral meson mixing and
decay width differences in partially-quenched, finite volume heavy
meson chiral perturbation theory relevant for lattice computations.
For heavy-light neutral meson mixing, inclusion of operators 
beyond those in the Standard Model complicates the chiral extrapolation as
two LECs appear at leading-order rather than one for the Standard Model
operator.  The matrix elements relevant for lifetime ratios
and decay width differences have similarly complicated light-quark
mass and spatial volume dependencies. This arises due
to the fact the operators for these processes do not guarantee 
that $B{-}\bar{B}$ and $B^{*}{-}\bar{B}^{*}$ mixing amplitudes 
are proportional to each other in the heavy-quark limit.  Our 
results are useful for current and future lattice calculations of
these matrix elements, which are needed in high-precision tests
of the Standard Model and the search for new physics.

\acknowledgments
The authors thank M.~J.~Savage, S.~R.~Sharpe and M.~Wingate for useful
discussions.  This work is supported by the U.S. Department of Energy
via grant DE-FG03-97ER41014.
CJDL is grateful for the hospitality of DAMTP, University of Cambridge and
CTS, National Taiwan University.

\appendix

\section{Integrals and sums}
\label{sec:integrals-sums}

We have regularised ultra-violet divergences that appear in the
various loop integrals using dimensional regularisation, and
subtracted the term
\begin{equation}
\label{eq:lambdabar}
 \bar{\lambda} = \frac{2}{4 - d} - \gamma_{E} + {\mathrm{log}}(4\pi) + 1.
\end{equation}
The integrals appearing in the full QCD calculation are defined by 
\begin{eqnarray}
 I_{\bar{\lambda}}(m) &\equiv&
 \mu^{4 - d}\int \frac{d^{d}k}{(2\pi)^{d}} \frac{1}{k^{2}-m^{2}+ i\epsilon}
\nonumber\\
   &=& \frac{i m^{2}}{16 \pi^{2}}
    \left [ \bar{\lambda} - {\log\left ( \frac{m^{2}}{\mu^{2}}\right )}
     \right ] , 
\end{eqnarray}
\begin{eqnarray}
 H_{\bar{\lambda}}(m,\Delta) &\equiv&
 \left ( g^{\rho\nu} - v^{\rho} v^{\nu}\right ) \mu^{4 - d}  
\nonumber\\&&\times
\frac{\partial}{\partial\Delta}\int
 \frac{d^{d}k}{(2\pi)^{d}} 
  \frac{k_{\rho} k_{\nu}}{(k^{2}-m^{2}
+ i\epsilon)(v\cdot k - \Delta + i\epsilon)}
\nonumber\\
 &=& 3 \frac{\partial}{\partial\Delta} F_{\bar{\lambda}}(m,\Delta),
\end{eqnarray}
where
%
%\begin{widetext}
\begin{equation}
\label{eq:FandR}
 F_{\bar{\lambda}}(m,\Delta) = \frac{i}{16\pi^{2}} \bigg \{ 
 \left [ \bar{\lambda} - {\log\left ( \frac{m^{2}}{\mu^{2}}\right )}
     \right ] \left ( \frac{2\Delta^{2}}{3} - m^{2}\right )\Delta
   + \left ( \frac{10\Delta^{2}}{9} - \frac{4 m^{2}}{3}\right ) \Delta
   + \frac{2 (\Delta^{2}-m^{2})}{3} m R\left ( \frac{\Delta}{m}\right )
\bigg \} ,
\end{equation}
%\end{widetext}
%
with
\begin{equation}
\label{eq:FandR_again}
R(x) \equiv \sqrt{x^{2}-1}\mbox{ }{\mathrm{log}}\left ( 
 \frac{x - \sqrt{x^{2}-1+i\epsilon}}{x + \sqrt{x^{2}-1+i\epsilon}} 
\right ) ,
\end{equation}
and $\mu$ is the renormalisation scale.  For the partially quenched
calculations, we also need the integrals
\begin{equation}
 I^{(\eta^{\prime})}_{\bar{\lambda}} \equiv
 \mu^{4 - d}\int \frac{d^{d}k}{(2\pi)^{d}} 
\frac{1}{(k^{2}-m^{2}+ i\epsilon)^{2}}
   = \frac{\partial I_{\bar{\lambda}}(m)}{\partial m^{2}} , 
\end{equation}
and
\begin{eqnarray}
 H^{\eta^{\prime}}_{\bar{\lambda}}(m,\Delta) 
&\equiv& 
 \left ( g^{\rho\nu} - v^{\rho} v^{\nu}\right ) \mu^{4 - d} 
\nonumber\\
&&\times
 \frac{\partial}{\partial\Delta}\int
 \frac{d^{d}k}{(2\pi)^{d}} 
  \frac{k_{\rho} k_{\nu}}{(k^{2}-m^{2}+i\epsilon)^{2}
  (v\cdot k - \Delta + i\epsilon)}
\nonumber\\
 &=& \frac{\partial}{\partial m^{2}} H_{\bar{\lambda}}(m,\Delta).
\end{eqnarray}
In a cubic spatial box of side length $L$ with periodic boundary
condition, the three-momenta are quantised as
\begin{equation}
\label{eq:k_quantum}
\vec{k} = \left (\frac{2\pi}{L}\right ) \vec{i} ,
\end{equation} 
and one instead obtains the sums (after subtracting the ultra-violet
divergences)
\begin{equation}
\label{eq:FV_I}
 {\mathcal{I}}(m) 
\equiv \frac{1}{L^{3}}\sum_{\vec{k}}\int \frac{d k_{0}}{2\pi}
  \frac{1}{k^{2}-m^{2}+i\epsilon} = I(m) + I_{{\mathrm{FV}}}(m),
\end{equation}
and
%
%\begin{widetext}
\begin{equation}
 {\mathcal{H}}(m,\Delta) \equiv
 \left ( g^{\rho\nu} - v^{\rho} v^{\nu}\right )\left ( \frac{1}{L^{3}}\right ) 
 \sum_{\vec{k}} 
\frac{\partial}{\partial\Delta}\int
 \frac{d k_{0}}{2\pi} 
  \frac{k_{\rho} k_{\nu}}{(k^{2}-m^{2}
+i\epsilon)(v\cdot k - \Delta + i\epsilon)}
\label{eq:FV_H}
 = H(m,\Delta) + H_{{\mathrm{FV}}}(m,\Delta)
\end{equation}
for the full QCD calculation, where
\begin{equation}
 I(m) = I_{\bar{\lambda}}(m)|_{\bar{\lambda}=0} ,
\end{equation}
and
\begin{equation}
 H(m) = H_{\bar{\lambda}}(m,\Delta)|_{\bar{\lambda}=0} ,
\end{equation}
are the infinite volume limits of ${\mathcal{I}}$ and ${\mathcal{H}}$,
and ($n=|\vec{n}|$)
\begin{eqnarray}
 I_{\mathrm{FV}}(m) 
  &=& \frac{-i}{4 \pi^{2}} m \sum_{\vec{n}\not=\vec{0}}
 \frac{1}{n L} K_{1}\left (n m L\right )\nonumber\\
& &\nonumber\\
 &\stackrel{m L \gg 1}{\longrightarrow}&
 \frac{-i}{4\pi^{2}} \sum_{\vec{n}\not=\vec{0}}
 \sqrt{\frac{m\pi}{2 n L}}
  \left (\frac{1}{n L}\right ) {\mathrm e}^{-n m L}
  \times \left \{
  1 + \frac{3}{8 n m L} 
 - \frac{15}{128 (n m L)^{2}}
 + \op\left ( \left [\frac{1}{n m L}\right ]^{3}\right )
 \right \}\,,
\end{eqnarray}
is the finite volume correction to $I(m)$.  The function $H_{\mathrm{FV}}$
is the finite volume correction to $H(m,\Delta)$ and can be obtained via
\begin{equation}
 H_{\mathrm{FV}}(m,\Delta) = i \bigg[ 
 (m^{2} - \Delta^{2}) K_{\mathrm{FV}}(m,\Delta) 
  - 2 \Delta J_{\mathrm{FV}}(m,\Delta)
  + i I_{\mathrm{FV}}(m)
 \bigg ] ,
\end{equation}
where $J_{\mathrm{FV}}(m,\Delta)$ and $K_{\mathrm{FV}}(m,\Delta)$ are
given by $ \left[w_{\vec{k}} = \sqrt{|\vec{k}|^{2}+m^{2}}\right]$
 \begin{equation}
\label{eq:JFV}
 J_{\mathrm{FV}}(m,\Delta) =
 \left (\frac{1}{2\pi} \right )^{2} 
 \sum_{\vec{n}\not = \vec{0}} \int_{0}^{\infty}
 d |\vec{k}| \left ( \frac{|\vec{k}|}{w_{\vec{k}}\mbox{ }
 (w_{\vec{k}} + \Delta)} \right )
 \left ( \frac{{\mathrm{sin}}(|\vec{k}| |\vec{n}| L)}{|\vec{n}| L}\right ) \,,
\end{equation}
and
\begin{equation}
\label{eq:KFV_asymptotic}
 K_{\mathrm{FV}}(m,\Delta) = 
 \frac{\partial J_{\mathrm{FV}}(m,\Delta)}
 {\partial \Delta}\,.
\end{equation}

In the asymptotic limit where $mL\gg 1$ it can be shown that 
(with $n\equiv |\vec{n}|$)
\begin{eqnarray}
\label{eq:JFV_asymp}
 J_{\mathrm{FV}}(m,\Delta) &=& 
\sum_{\vec{n}\not=\vec{0}} \left ( \frac{1}{8\pi n L}\right )
{\mathrm{e}}^{- n m L} 
{\mathcal{A}} ,
\end{eqnarray}
where
\begin{eqnarray}
 {\mathcal{A}} &=&{\mathrm e}^{(z^{2})} \big [ 
1 - {\mathrm{Erf}}(z)\mbox{ }\big ]
+\left (\frac{1}{n m L} \right ) \bigg [
 \frac{1}{\sqrt{\pi}} \left ( \frac{z}{4} - 
\frac{z^{3}}{2}\right )
 + \frac{z^{4}}{2}{\mathrm e}^{(z^{2})} 
 \big [ 1 - {\mathrm{Erf}}(z)\mbox{ }\big ]
\bigg ]\nonumber\\
\label{eq:JFV_asymptotic}
& &
-\left (\frac{1}{n m L} \right )^{2}\bigg [
\frac{1}{\sqrt{\pi}}\left ( \frac{9z}{64} - 
\frac{5z^{3}}{32}
  +\frac{7z^{5}}{16} + \frac{z^{7}}{8} \right )
-\left ( \frac{z^{6}}{2} + \frac{z^{8}}{8}\right )
{\mathrm e}^{(z^{2})} \big [ 1 - {\mathrm{Erf}}(z)\mbox{ }\big ]
\bigg ]
+\op\left (\left [ \frac{1}{n m L}\right ]^{3}\right ) ,
\end{eqnarray}
with
\begin{equation}
z \equiv \left (\frac{\Delta}{m}\right ) 
\sqrt{\frac{n m L}{2}} .
\end{equation}
The quantity ${\mathcal{A}}$ is the alteration of finite volume
effects due to the presence of a non-zero $\Delta$. See
Ref.~\cite{Arndt:2004bg} for further discussion.

For the PQ$\chi$PT calculations, one also needs
\begin{equation}
\label{eq:FV_I_DP}
 {\mathcal{I}}^{\eta^{\prime}}(m) \equiv \frac{1}{L^{3}}
   \sum_{\vec{k}}\int \frac{d k_{0}}{2\pi}
  \frac{1}{(k^{2}-m^{2}+i\epsilon)^{2}} = \frac{\partial I(m)}{\partial m^{2}}
 + \frac{\partial I_{\mathrm{FV}}(m)}{\partial m^{2}} \,,
\end{equation}
and
\begin{eqnarray}
 {\mathcal{H}}^{\eta^{\prime}}(m,\Delta) 
&\equiv& \frac{\partial}{\partial\Delta}
 \left [
 \left ( g^{\rho\nu} - v^{\rho} v^{\nu}\right )\left ( \frac{1}{L^{3}}\right ) 
 \sum_{\vec{k}} \int
 \frac{d k_{0}}{2\pi} 
  \frac{k_{\rho} k_{\nu}}{(k^{2}-m^{2}
+i\epsilon)^{2}(v\cdot k - \Delta + i\epsilon)}
 \right ] \nonumber\\
& &\nonumber\\
\label{eq:FV_H_DP}
 &=& \frac{\partial H(m,\Delta)}{\partial m^{2}} + 
  \frac{\partial H_{\mathrm{FV}}(m,\Delta)}{\partial m^{2}} \,.
\end{eqnarray}

\section{Loop contributions and bag parameters}
\label{sec:loop-contributions}

In this appendix, we present results for the various contributions in
Eqs.~(\ref{eq:7}) and (\ref{eq:10}), ${\cal W}_{B^{0}_{(s)}}$, ${\cal
  W}_{\bar{B}^{0}_{(s)}}$, ${\cal T}^{(i)}_{d(s)}$ and ${\cal
  Q}^{(i)}_{d(s)}$ ($i=1,2,3,4,5$).  These results are given in the
sea and valence isospin limit of SU(6$|$3)
partially-quenched HM$\chi$PT, with the quark masses given in
Eq.~(\ref{eq:PQ_mass_matrix}).  The QCD limit, where sea and valence
quark masses are equal, is easily taken by setting
$m_{j}=m_{u}$ and $m_{r}=m_{s}$. We also present the bag parameters
defined in Eqs.~(\ref{eq:8}) and (\ref{eq:9}).
\subsection{Loop contributions in SU(6$|$3) partially-quenched heavy meson
  chiral perturbation theory}

To compactly express the partially quenched expressions, it is useful
to define the following quantities:
\begin{eqnarray}
  \label{eq:11}
  A_{u,u} &=& \frac{2 \left(\delta_{VS}^2-M_{\pi }^2+M_X^2\right)
 \delta_{VS}^2}{\left(M_{\pi }^2-M_X^2\right)^2}+\frac{3}{2}\,,
\end{eqnarray}
\begin{eqnarray}
  \label{eq:12}
  B_{u,u} &=& 1- A_{u,u}\,,
\end{eqnarray}
\begin{eqnarray}
  \label{eq:13}
  C_{u,u} &=& 3 \delta_{VS}^2-\frac{2 \delta_{VS}^4}{M_{\pi }^2-M_X^2}\,,
\end{eqnarray}
\begin{eqnarray}
  \label{eq:14}
  A_{s,s} &=& \frac{3 \left(8 \delta_{VSs}^4+\left(2 \delta_{VS}
        ^2-M_{\pi }^2+M_{s,s}^2\right)^2\right)}{\left(2 \delta_{VS}^2+4
      \delta_{VSs}^2-M_{\pi }^2+M_{s,s}^2\right)^2} \,,
\end{eqnarray}
\begin{eqnarray}
  \label{eq:15}
  B_{s,s} &=& 1-A_{s,s}\,,
\end{eqnarray}
\begin{eqnarray}
  \label{eq:16}
  C_{s,s} &=& \frac{6 \delta_{VSs}^2 \left(2 \delta_{VS} ^2-M_{\pi
      }^2+M_{s,s}^2\right)}{\left(2 \delta_{VS}^2+4
      \delta_{VSs}^2-M_{\pi}^2+M_{s,s}^2\right)^2}\,, 
\end{eqnarray}
with $M^{2}_{a,b} = B_{0}(m_{a} + m_{b})$,
$\delta_{VS}^2=M_\pi^2-M_{u,j}^2$,
$\delta_{VSs}^2=M_{s,s}^2-M_{s,r}^2$, $M_\pi=M_{u,u}$, and
$M_X^2=\frac{1}{3}(M_\pi^2+2M_{s,s}^2-2\delta_{VS}^2-4\delta_{VSs}^2)$.
For ease of use, we note that in the QCD limit (setting valence and
sea masses to be identical),
\begin{displaymath}
  A_{u,u}^{QCD} = \frac{3}{2},\quad\quad
    B_{u,u}^{QCD} = -\frac{1}{2},\quad\quad
C_{u,u}^{QCD}=0,
\end{displaymath}
\begin{displaymath}
  A_{s,s}^{QCD} = 3,\quad\quad
    B_{s,s}^{QCD} =-2,\quad\quad
C_{s,s}^{QCD}=0.
\end{displaymath}

The various loop contributions can then be written as
\begin{eqnarray}
  {\cal W}_{B^0} = {\cal W}_{\bar{B}^0} &=& -\frac{i\, g^2}{3
   f^2} \Big[B_{u,u} {\cal H}\left(M_X,\Delta_{\ast}\right)
    -6 {\cal H}\left(M_{u,j},\Delta_{\ast}+\delta
   _{{uj}}\right)-3 {\cal H}\left(M_{u,r},\Delta_{\ast}
     +\delta _{{sr}}+\delta _{{us}}\right)
\nonumber
\\
&& \hspace*{6cm}
+A_{u,u}
   {\cal H}\left(M_{u,u},\Delta_{\ast}\right)
+C_{u,u} {\cal H}^{\eta
  '}\left(M_{u,u},\Delta_{\ast}\right)\Big] \,,
  \label{eq:17}
%%%%%%%%%%%%%%%%%
\\
%%%%%%%%%%%%%%%%%
  {\cal W}_{B^0_s} = {\cal W}_{\bar{B}^0_s} &=& 
%%%%%%%%%%%%%%%%%
-\frac{i\, g^2}{3
   f^2} \Big[B_{s,s} {\cal H}\left(M_X,\Delta_{\ast}\right)-6 {\cal
   H}\left(M_{s,j},\Delta_{\ast}+\delta 
   _{{uj}}-\delta _{{us}}\right)-3 {\cal
   H}\left(M_{s,r},\Delta_{\ast}+\delta _{{sr}}\right) 
\nonumber
\\
&& \hspace*{6cm}
+A_{s,s}
   {\cal H}\left(M_{s,s},\Delta_{\ast}\right)
+C_{s,s} {\cal H}^{\eta
  '}\left(M_{s,s},\Delta_{\ast}\right)\Big]
\,,
\end{eqnarray}
for the wave-function renormalisations,
\begin{eqnarray}
  \label{eq:18}
  {\cal T}_{d}^{(1,2,3)} &=&
\frac{i}{3 f^2} \left[2 B_{u,u} {\cal I}\left(M_X\right)
-6 {\cal I}\left(M_{u,j}\right)
-3 {\cal I}\left(M_{u,r}\right)
+\left(2 A_{u,u}-3\right){\cal I}\left(M_{u,u}\right)
+2 C_{u,u} {\cal I}^{\eta'}\left(M_{u,u}\right)\right] \,,
\\
 {\cal T}_{s}^{(1,2,3)}  &=&
\frac{i}{3 f^2} \left[2 B_{s,s} {\cal I}\left(M_X\right)
-6 {\cal I}\left(M_{s,j}\right)
-3 {\cal I}\left(M_{s,r}\right)
+\left(2 A_{s,s}-3\right){\cal I}\left(M_{s,s}\right)
+2 C_{s,s} {\cal I}^{\eta'}\left(M_{s,s}\right)\right] \,,
\\
  {\cal T}_{d}^{(4,5)}  &=&
-\frac{i}{f^2} \left[
    2{\cal I}\left(M_{u,j}\right)+ 
   {\cal I}\left(M_{u,r}\right)-
   {\cal I}\left(M_{u,u}\right)\right]\,,
\\
  {\cal T}_{s}^{(4,5)} &=&
-\frac{i}{f^2} \left[
    2{\cal I}\left(M_{s,j}\right)+ 
   {\cal I}\left(M_{s,r}\right)-
   {\cal I}\left(M_{s,s}\right)\right]\,,
\end{eqnarray}
for tadpole integrals [Fig \ref{fig:oneloop}(b)] and,
\begin{eqnarray}
  \label{eq:19}
 {\cal Q}^{(i)}_{d} &=&
\frac{i g^{2}}{3 f^2} \left(B_{u,u} {\cal H}\left(M_X,\Delta
  \right)+\left(A_{u,u}-3\right){\cal H}\left(M_{u,u},\Delta \right)
+C_{u,u} {\cal H}^{\eta'}\left[M_{u,u},\Delta \right]\right)\,,
\\
  {\cal Q}^{(i)}_{s} &=&
\frac{i g^{2}}{3 f^2} \left(B_{s,s} {\cal H}\left(M_X,\Delta
  \right)+\left(A_{s,s}-3\right) {\cal H}\left(M_{s,s},\Delta \right)
+C_{s,s} {\cal H}^{\eta'}\left[M_{s,s},\Delta \right]\right)\,,
\end{eqnarray}
for ``sunset'' integrals [Fig \ref{fig:oneloop}(c)].

\subsection{Bag parameters}
\label{sec:B-parameters}

For completeness, the bag parameters defined in Eqs.~(\ref{eq:8}) and
(\ref{eq:9}) are given by:
\begin{eqnarray}
  \label{eq:20}
  B^{(1)}_{B_d}(\mu) &=&
\frac{3 \beta _1}{8 \kappa ^2} \left(1+\frac{X_{I,u}}{f^2} + 
\frac{g^2 X_{H,u} }{f^2}\right)
\,,\\
  B^{(1)}_{B_s}(\mu) &=&
\frac{3 \beta _1}{8 \kappa ^2} \left(1+\frac{X_{I,s}}{f^2} + 
\frac{g^2 X_{H,s} }{f^2}\right)
\,,\\
  B^{({2/3})}_{B_d}(\mu) &=&
\frac{\beta _{2/3}}{\kappa ^2 \eta_{2/3} R^2}  \left(
1 + \frac{X_{I,u}}{f^2} + \frac{ g^2 \beta^\prime_{2/3}
    X_{H,u}}{f^2 \beta_{2/3}}\right)
\,,\\
  B^{({2/3})}_{B_s}(\mu) &=&
\frac{\beta _{2/3}}{\kappa ^2\eta_{2/3}R^2}  \left(
1 + \frac{X_{I,s}}{f^2} + \frac{ g^2 \beta^\prime_{2/3}
    X_{H,s}}{f^2 \beta_{2/3}}\right)
\,,\\
  B^{({4/5})}_{B_d}(\mu) &=& 
\frac{\beta_{4/5}+\hat{\beta }_{4/5}}{\kappa ^2 \eta_{4/5}R^2} 
    \left(1+ \frac{X_{I,u}}{f^2}+
      \frac{ g^2\left(\beta^\prime_{4/5}+\hat{\beta}^\prime_{4/5}\right)  
   X_{H,u} }{f^2 \left(\beta _{4/5}+\hat{\beta
     }_{4/5}\right)}\right) 
\,,\\
  B^{({4/5})}_{B_s}(\mu) &=&
\frac{\beta_{4/5}+\hat{\beta }_{4/5}}{\kappa ^2 \eta_{4/5}R^2} 
    \left(1+ \frac{X_{I,s}}{f^2}+
      \frac{ g^2\left(\beta^\prime_{4/5}+\hat{\beta}^\prime_{4/5}\right)  
   X_{H,s} }{f^2 \left(\beta _{4/5}+\hat{\beta
     }_{4/5}\right)}\right) 
\,,
\end{eqnarray}
where $\kappa$ is the LEC governing the heavy-light axial current
\cite{Arndt:2004bg} and for convenience we have defined
\begin{eqnarray}
  \label{eq:21}
  X_{H,u}&=&\frac{i}{3}\left(B_{u,u}{\cal H}\left(M_X,\Delta
    \right)
+\left(A_{u,u}-3\right){\cal H}\left(M_{u,u},\Delta \right)
+C_{u,u} {\cal H}^{\eta'}\left[M_{u,u},\Delta \right]\right)\,,
\\
X_{H,s} &=&\frac{i}{3} \left(B_{s,s} {\cal H}\left(M_X,\Delta \right)
+\left(A_{s,s}-3\right) {\cal H}\left(M_{s,s},\Delta \right)
+C_{s,s} {\cal H}^{\eta'}\left[M_{s,s},\Delta \right]\right) \,,
\\
X_{I,u} &=&\frac{i}{3}  \left(B_{u,u} {\cal I}\left(M_X\right)
+\left(A_{u,u}-3\right) {\cal I}\left(M_{u,u}\right)
+C_{u,u} {\cal I}^{\eta'}\left[M_{u,u}\right]\right) \,,
\\
X_{I,s} &=& \frac{i}{3}  \left(B_{s,s} {\cal I}\left(M_X\right)
+\left(A_{s,s}-3\right) {\cal I}\left(M_{s,s}\right)
+C_{s,s} {\cal I}^{\eta'}\left[M_{s,s}\right]\right) \,.
\end{eqnarray}

%%%%%%%%%%%%%%%%%%%%%%%%%%%%%%%%%%%%%%%%%%%%%%%%%%%%%%%%%
%                           BIBLIOGRAPHY
%%%%%%%%%%%%%%%%%%%%%%%%%%%%%%%%%%%%%%%%%%%%%%%%%%%%%%%%%%

\bibliography{refs}

\begin{thebibliography}{38}
\expandafter\ifx\csname natexlab\endcsname\relax\def\natexlab#1{#1}\fi
\expandafter\ifx\csname bibnamefont\endcsname\relax
  \def\bibnamefont#1{#1}\fi
\expandafter\ifx\csname bibfnamefont\endcsname\relax
  \def\bibfnamefont#1{#1}\fi
\expandafter\ifx\csname citenamefont\endcsname\relax
  \def\citenamefont#1{#1}\fi
\expandafter\ifx\csname url\endcsname\relax
  \def\url#1{\texttt{#1}}\fi
\expandafter\ifx\csname urlprefix\endcsname\relax\def\urlprefix{URL }\fi
\providecommand{\bibinfo}[2]{#2}
\providecommand{\eprint}[2][]{\url{#2}}

\bibitem[{\citenamefont{Abulencia et~al.}(2006)}]{Abulencia:2006ze}
\bibinfo{author}{\bibfnamefont{A.}~\bibnamefont{Abulencia}}
  \bibnamefont{et~al.} (\bibinfo{collaboration}{CDF}) (\bibinfo{year}{2006}),
  \eprint{hep-ex/0609040}.

\bibitem[{\citenamefont{Bergmann et~al.}(2000)\citenamefont{Bergmann, Grossman,
  Ligeti, Nir, and Petrov}}]{Bergmann:2000id}
\bibinfo{author}{\bibfnamefont{S.}~\bibnamefont{Bergmann}},
  \bibinfo{author}{\bibfnamefont{Y.}~\bibnamefont{Grossman}},
  \bibinfo{author}{\bibfnamefont{Z.}~\bibnamefont{Ligeti}},
  \bibinfo{author}{\bibfnamefont{Y.}~\bibnamefont{Nir}}, \bibnamefont{and}
  \bibinfo{author}{\bibfnamefont{A.~A.} \bibnamefont{Petrov}},
  \bibinfo{journal}{Phys. Lett.} \textbf{\bibinfo{volume}{B486}},
  \bibinfo{pages}{418} (\bibinfo{year}{2000}), \eprint{hep-ph/0005181}.

\bibitem[{\citenamefont{Gabbiani et~al.}(1996)\citenamefont{Gabbiani,
  Gabrielli, Masiero, and Silvestrini}}]{Gabbiani:1996hi}
\bibinfo{author}{\bibfnamefont{F.}~\bibnamefont{Gabbiani}},
  \bibinfo{author}{\bibfnamefont{E.}~\bibnamefont{Gabrielli}},
  \bibinfo{author}{\bibfnamefont{A.}~\bibnamefont{Masiero}}, \bibnamefont{and}
  \bibinfo{author}{\bibfnamefont{L.}~\bibnamefont{Silvestrini}},
  \bibinfo{journal}{Nucl. Phys.} \textbf{\bibinfo{volume}{B477}},
  \bibinfo{pages}{321} (\bibinfo{year}{1996}), \eprint{hep-ph/9604387}.

\bibitem[{\citenamefont{Yao et~al.}(2006)}]{Yao:2006px}
\bibinfo{author}{\bibfnamefont{W.~M.} \bibnamefont{Yao}} \bibnamefont{et~al.}
  (\bibinfo{collaboration}{Particle Data Group}), \bibinfo{journal}{J. Phys.}
  \textbf{\bibinfo{volume}{G33}}, \bibinfo{pages}{1} (\bibinfo{year}{2006}).

\bibitem[{\citenamefont{Beneke et~al.}(1996)\citenamefont{Beneke, Buchalla, and
  Dunietz}}]{Beneke:1996gn}
\bibinfo{author}{\bibfnamefont{M.}~\bibnamefont{Beneke}},
  \bibinfo{author}{\bibfnamefont{G.}~\bibnamefont{Buchalla}}, \bibnamefont{and}
  \bibinfo{author}{\bibfnamefont{I.}~\bibnamefont{Dunietz}},
  \bibinfo{journal}{Phys. Rev.} \textbf{\bibinfo{volume}{D54}},
  \bibinfo{pages}{4419} (\bibinfo{year}{1996}), \eprint{hep-ph/9605259}.

\bibitem[{\citenamefont{Beneke et~al.}(1999)\citenamefont{Beneke, Buchalla,
  Greub, Lenz, and Nierste}}]{Beneke:1998sy}
\bibinfo{author}{\bibfnamefont{M.}~\bibnamefont{Beneke}},
  \bibinfo{author}{\bibfnamefont{G.}~\bibnamefont{Buchalla}},
  \bibinfo{author}{\bibfnamefont{C.}~\bibnamefont{Greub}},
  \bibinfo{author}{\bibfnamefont{A.}~\bibnamefont{Lenz}}, \bibnamefont{and}
  \bibinfo{author}{\bibfnamefont{U.}~\bibnamefont{Nierste}},
  \bibinfo{journal}{Phys. Lett.} \textbf{\bibinfo{volume}{B459}},
  \bibinfo{pages}{631} (\bibinfo{year}{1999}), \eprint{hep-ph/9808385}.

\bibitem[{\citenamefont{Lenz and Nierste}(2006)}]{Lenz:2006hd}
\bibinfo{author}{\bibfnamefont{A.}~\bibnamefont{Lenz}} \bibnamefont{and}
  \bibinfo{author}{\bibfnamefont{U.}~\bibnamefont{Nierste}}
  (\bibinfo{year}{2006}), \eprint{hep-ph/0612167}.

\bibitem[{\citenamefont{Onogi}(2006)}]{Onogi:2006km}
\bibinfo{author}{\bibfnamefont{T.}~\bibnamefont{Onogi}},
  \bibinfo{journal}{PoS.} \textbf{\bibinfo{volume}{LAT2006}},
  \bibinfo{pages}{017} (\bibinfo{year}{2006}), \eprint{hep-lat/0610115}.

\bibitem[{\citenamefont{Burdman and Donoghue}(1992)}]{Burdman:1992gh}
\bibinfo{author}{\bibfnamefont{G.}~\bibnamefont{Burdman}} \bibnamefont{and}
  \bibinfo{author}{\bibfnamefont{J.~F.} \bibnamefont{Donoghue}},
  \bibinfo{journal}{Phys. Lett.} \textbf{\bibinfo{volume}{B280}},
  \bibinfo{pages}{287} (\bibinfo{year}{1992}).

\bibitem[{\citenamefont{Wise}(1992)}]{Wise:1992hn}
\bibinfo{author}{\bibfnamefont{M.~B.} \bibnamefont{Wise}},
  \bibinfo{journal}{Phys. Rev.} \textbf{\bibinfo{volume}{D45}},
  \bibinfo{pages}{2188} (\bibinfo{year}{1992}).

\bibitem[{\citenamefont{Yan et~al.}(1992)}]{Yan:1992gz}
\bibinfo{author}{\bibfnamefont{T.-M.} \bibnamefont{Yan}} \bibnamefont{et~al.},
  \bibinfo{journal}{Phys. Rev.} \textbf{\bibinfo{volume}{D46}},
  \bibinfo{pages}{1148} (\bibinfo{year}{1992}).

\bibitem[{\citenamefont{Arndt and Lin}(2004)}]{Arndt:2004bg}
\bibinfo{author}{\bibfnamefont{D.}~\bibnamefont{Arndt}} \bibnamefont{and}
  \bibinfo{author}{\bibfnamefont{C.~J.~D.} \bibnamefont{Lin}},
  \bibinfo{journal}{Phys. Rev.} \textbf{\bibinfo{volume}{D70}},
  \bibinfo{pages}{014503} (\bibinfo{year}{2004}), \eprint{hep-lat/0403012}.

\bibitem[{\citenamefont{Becirevic et~al.}(2006)\citenamefont{Becirevic, Fajfer,
  and Kamenik}}]{Becirevic:2006me}
\bibinfo{author}{\bibfnamefont{D.}~\bibnamefont{Becirevic}},
  \bibinfo{author}{\bibfnamefont{S.}~\bibnamefont{Fajfer}}, \bibnamefont{and}
  \bibinfo{author}{\bibfnamefont{J.}~\bibnamefont{Kamenik}}
  (\bibinfo{year}{2006}), \eprint{hep-ph/0612224}.

\bibitem[{\citenamefont{Sharpe and Shoresh}(2001)}]{Sharpe:2001fh}
\bibinfo{author}{\bibfnamefont{S.~R.} \bibnamefont{Sharpe}} \bibnamefont{and}
  \bibinfo{author}{\bibfnamefont{N.}~\bibnamefont{Shoresh}},
  \bibinfo{journal}{Phys. Rev.} \textbf{\bibinfo{volume}{D64}},
  \bibinfo{pages}{114510} (\bibinfo{year}{2001}), \eprint{hep-lat/0108003}.

\bibitem[{\citenamefont{Booth}(1995)}]{Booth:1995hx}
\bibinfo{author}{\bibfnamefont{M.~J.} \bibnamefont{Booth}},
  \bibinfo{journal}{Phys. Rev.} \textbf{\bibinfo{volume}{D51}},
  \bibinfo{pages}{2338} (\bibinfo{year}{1995}), \eprint{hep-ph/9411433}.

\bibitem[{\citenamefont{Sharpe and Zhang}(1996)}]{Sharpe:1996qp}
\bibinfo{author}{\bibfnamefont{S.~R.} \bibnamefont{Sharpe}} \bibnamefont{and}
  \bibinfo{author}{\bibfnamefont{Y.}~\bibnamefont{Zhang}},
  \bibinfo{journal}{Phys. Rev.} \textbf{\bibinfo{volume}{D53}},
  \bibinfo{pages}{5125} (\bibinfo{year}{1996}), \eprint{hep-lat/9510037}.

\bibitem[{\citenamefont{Boyd and Grinstein}(1995)}]{Boyd:1995pa}
\bibinfo{author}{\bibfnamefont{C.~G.} \bibnamefont{Boyd}} \bibnamefont{and}
  \bibinfo{author}{\bibfnamefont{B.}~\bibnamefont{Grinstein}},
  \bibinfo{journal}{Nucl. Phys.} \textbf{\bibinfo{volume}{B442}},
  \bibinfo{pages}{205} (\bibinfo{year}{1995}), \eprint{hep-ph/9402340}.

\bibitem[{\citenamefont{Booth}()}]{Booth:1994rr}
\bibinfo{author}{\bibfnamefont{M.~J.} \bibnamefont{Booth}}, \bibinfo{note}{{\bf
  hep-ph/9412228}}.

\bibitem[{\citenamefont{Bernard and Golterman}(1992)}]{Bernard:1992mk}
\bibinfo{author}{\bibfnamefont{C.~W.} \bibnamefont{Bernard}} \bibnamefont{and}
  \bibinfo{author}{\bibfnamefont{M.~F.~L.} \bibnamefont{Golterman}},
  \bibinfo{journal}{Phys. Rev.} \textbf{\bibinfo{volume}{D46}},
  \bibinfo{pages}{853} (\bibinfo{year}{1992}), \eprint{hep-lat/9204007}.

\bibitem[{\citenamefont{Bernard and Golterman}(1994)}]{Bernard:1994sv}
\bibinfo{author}{\bibfnamefont{C.~W.} \bibnamefont{Bernard}} \bibnamefont{and}
  \bibinfo{author}{\bibfnamefont{M.~F.~L.} \bibnamefont{Golterman}},
  \bibinfo{journal}{Phys. Rev.} \textbf{\bibinfo{volume}{D49}},
  \bibinfo{pages}{486} (\bibinfo{year}{1994}), \eprint{hep-lat/9306005}.

\bibitem[{\citenamefont{Sharpe and Shoresh}(2000)}]{Sharpe:2000bc}
\bibinfo{author}{\bibfnamefont{S.~R.} \bibnamefont{Sharpe}} \bibnamefont{and}
  \bibinfo{author}{\bibfnamefont{N.}~\bibnamefont{Shoresh}},
  \bibinfo{journal}{Phys. Rev.} \textbf{\bibinfo{volume}{D62}},
  \bibinfo{pages}{094503} (\bibinfo{year}{2000}), \eprint{hep-lat/0006017}.

\bibitem[{\citenamefont{Grinstein et~al.}(1992)\citenamefont{Grinstein,
  Jenkins, Manohar, Savage, and Wise}}]{Grinstein:1992qt}
\bibinfo{author}{\bibfnamefont{B.}~\bibnamefont{Grinstein}},
  \bibinfo{author}{\bibfnamefont{E.}~\bibnamefont{Jenkins}},
  \bibinfo{author}{\bibfnamefont{A.~V.} \bibnamefont{Manohar}},
  \bibinfo{author}{\bibfnamefont{M.~J.} \bibnamefont{Savage}},
  \bibnamefont{and} \bibinfo{author}{\bibfnamefont{M.~B.} \bibnamefont{Wise}},
  \bibinfo{journal}{Nucl. Phys.} \textbf{\bibinfo{volume}{B380}},
  \bibinfo{pages}{369} (\bibinfo{year}{1992}), \eprint{hep-ph/9204207}.

\bibitem[{\citenamefont{Becirevic and Villadoro}(2004)}]{Becirevic:2004qd}
\bibinfo{author}{\bibfnamefont{D.}~\bibnamefont{Becirevic}} \bibnamefont{and}
  \bibinfo{author}{\bibfnamefont{G.}~\bibnamefont{Villadoro}},
  \bibinfo{journal}{Phys. Rev.} \textbf{\bibinfo{volume}{D70}},
  \bibinfo{pages}{094036} (\bibinfo{year}{2004}), \eprint{hep-lat/0408029}.

\bibitem[{\citenamefont{Isgur and Wise}(1990)}]{Isgur:1989ed}
\bibinfo{author}{\bibfnamefont{N.}~\bibnamefont{Isgur}} \bibnamefont{and}
  \bibinfo{author}{\bibfnamefont{M.~B.} \bibnamefont{Wise}},
  \bibinfo{journal}{Phys. Lett.} \textbf{\bibinfo{volume}{B237}},
  \bibinfo{pages}{527} (\bibinfo{year}{1990}).

\bibitem[{\citenamefont{Isgur and Wise}(1989)}]{Isgur:1989vq}
\bibinfo{author}{\bibfnamefont{N.}~\bibnamefont{Isgur}} \bibnamefont{and}
  \bibinfo{author}{\bibfnamefont{M.~B.} \bibnamefont{Wise}},
  \bibinfo{journal}{Phys. Lett.} \textbf{\bibinfo{volume}{B232}},
  \bibinfo{pages}{113} (\bibinfo{year}{1989}).

\bibitem[{\citenamefont{Grinstein}(1990)}]{Grinstein:1990mj}
\bibinfo{author}{\bibfnamefont{B.}~\bibnamefont{Grinstein}},
  \bibinfo{journal}{Nucl. Phys.} \textbf{\bibinfo{volume}{B339}},
  \bibinfo{pages}{253} (\bibinfo{year}{1990}).

\bibitem[{\citenamefont{Eichten and Hill}(1990)}]{Eichten:1989zv}
\bibinfo{author}{\bibfnamefont{E.}~\bibnamefont{Eichten}} \bibnamefont{and}
  \bibinfo{author}{\bibfnamefont{B.}~\bibnamefont{Hill}},
  \bibinfo{journal}{Phys. Lett.} \textbf{\bibinfo{volume}{B234}},
  \bibinfo{pages}{511} (\bibinfo{year}{1990}).

\bibitem[{\citenamefont{Georgi}(1990)}]{Georgi:1990um}
\bibinfo{author}{\bibfnamefont{H.}~\bibnamefont{Georgi}},
  \bibinfo{journal}{Phys. Lett.} \textbf{\bibinfo{volume}{B240}},
  \bibinfo{pages}{447} (\bibinfo{year}{1990}).

\bibitem[{\citenamefont{Flynn et~al.}(1991)\citenamefont{Flynn, Hernandez, and
  Hill}}]{Flynn:1990qz}
\bibinfo{author}{\bibfnamefont{J.~M.} \bibnamefont{Flynn}},
  \bibinfo{author}{\bibfnamefont{O.~F.} \bibnamefont{Hernandez}},
  \bibnamefont{and} \bibinfo{author}{\bibfnamefont{B.~R.} \bibnamefont{Hill}},
  \bibinfo{journal}{Phys. Rev.} \textbf{\bibinfo{volume}{D43}},
  \bibinfo{pages}{3709} (\bibinfo{year}{1991}).

\bibitem[{\citenamefont{Savage and Wise}(1990)}]{Savage:1990di}
\bibinfo{author}{\bibfnamefont{M.~J.} \bibnamefont{Savage}} \bibnamefont{and}
  \bibinfo{author}{\bibfnamefont{M.~B.} \bibnamefont{Wise}},
  \bibinfo{journal}{Phys. Lett.} \textbf{\bibinfo{volume}{B248}},
  \bibinfo{pages}{177} (\bibinfo{year}{1990}).

\bibitem[{\citenamefont{Becirevic et~al.}(2002)\citenamefont{Becirevic,
  Gimenez, Martinelli, Papinutto, and Reyes}}]{Becirevic:2001xt}
\bibinfo{author}{\bibfnamefont{D.}~\bibnamefont{Becirevic}},
  \bibinfo{author}{\bibfnamefont{V.}~\bibnamefont{Gimenez}},
  \bibinfo{author}{\bibfnamefont{G.}~\bibnamefont{Martinelli}},
  \bibinfo{author}{\bibfnamefont{M.}~\bibnamefont{Papinutto}},
  \bibnamefont{and} \bibinfo{author}{\bibfnamefont{J.}~\bibnamefont{Reyes}},
  \bibinfo{journal}{JHEP} \textbf{\bibinfo{volume}{04}}, \bibinfo{pages}{025}
  (\bibinfo{year}{2002}), \eprint{hep-lat/0110091}.

\bibitem[{\citenamefont{Gimenez and Reyes}(2001)}]{Gimenez:2000jj}
\bibinfo{author}{\bibfnamefont{V.}~\bibnamefont{Gimenez}} \bibnamefont{and}
  \bibinfo{author}{\bibfnamefont{J.}~\bibnamefont{Reyes}},
  \bibinfo{journal}{Nucl. Phys. Proc. Suppl.} \textbf{\bibinfo{volume}{94}},
  \bibinfo{pages}{350} (\bibinfo{year}{2001}), \eprint{hep-lat/0010048}.

\bibitem[{\citenamefont{Hashimoto and Yamada}(2001)}]{Hashimoto:2001zq}
\bibinfo{author}{\bibfnamefont{S.}~\bibnamefont{Hashimoto}} \bibnamefont{and}
  \bibinfo{author}{\bibfnamefont{N.}~\bibnamefont{Yamada}}
  (\bibinfo{collaboration}{JLQCD}) (\bibinfo{year}{2001}),
  \eprint{hep-ph/0104080}.

\bibitem[{\citenamefont{Dalgic et~al.}(2006)}]{Dalgic:2006gp}
\bibinfo{author}{\bibfnamefont{E.}~\bibnamefont{Dalgic}} \bibnamefont{et~al.}
  (\bibinfo{year}{2006}), \eprint{hep-lat/0610104}.

\bibitem[{\citenamefont{Ahmed et~al.}(2001)}]{Ahmed:2001xc}
\bibinfo{author}{\bibfnamefont{S.}~\bibnamefont{Ahmed}} \bibnamefont{et~al.}
  (\bibinfo{collaboration}{CLEO}), \bibinfo{journal}{Phys. Rev. Lett.}
  \textbf{\bibinfo{volume}{87}}, \bibinfo{pages}{251801}
  (\bibinfo{year}{2001}), \eprint{hep-ex/0108013}.

\bibitem[{\citenamefont{Anastassov et~al.}(2002)}]{Anastassov:2001cw}
\bibinfo{author}{\bibfnamefont{A.}~\bibnamefont{Anastassov}}
  \bibnamefont{et~al.} (\bibinfo{collaboration}{CLEO}), \bibinfo{journal}{Phys.
  Rev.} \textbf{\bibinfo{volume}{D65}}, \bibinfo{pages}{032003}
  (\bibinfo{year}{2002}), \eprint{hep-ex/0108043}.

\bibitem[{\citenamefont{Ginsparg and Wilson}(1982)}]{Ginsparg:1981bj}
\bibinfo{author}{\bibfnamefont{P.~H.} \bibnamefont{Ginsparg}} \bibnamefont{and}
  \bibinfo{author}{\bibfnamefont{K.~G.} \bibnamefont{Wilson}},
  \bibinfo{journal}{Phys. Rev.} \textbf{\bibinfo{volume}{D25}},
  \bibinfo{pages}{2649} (\bibinfo{year}{1982}).

\bibitem[{\citenamefont{Chen et~al.}(2006)\citenamefont{Chen, O'Connell, and
  Walker-Loud}}]{Chen:2006wf}
\bibinfo{author}{\bibfnamefont{J.-W.} \bibnamefont{Chen}},
  \bibinfo{author}{\bibfnamefont{D.}~\bibnamefont{O'Connell}},
  \bibnamefont{and}
  \bibinfo{author}{\bibfnamefont{A.}~\bibnamefont{Walker-Loud}}
  (\bibinfo{year}{2006}), \eprint{hep-lat/0611003}.

\end{thebibliography}
 
\end{document}